# The development of HISPEC for Keck and MODHIS for TMT: science cases and predicted sensitivities

Quinn M. Konopacky [a] [*], Ashley D. Baker[b], Dimitri Mawet[b,c], Michael P. Fitzgerald[d], Nemanja Jovanovic[b], Charles Beichman[e], Garreth Ruane[b,c], Rob Bertz[b], Hiroshi Terada[f], Richard Dekany[b], Larry Lingvay[b], Marc Kassis[g], David Anderson[f], Motohide Tamura[h,i], Björn Benneke[j], Thomas Beatty[k], Tuan Do[d], Shogo Nishiyama[l], Peter Plavchan[m], Jason Wang[n], Ji Wang[o], Adam Burgasser[a], Jean-Baptiste Ruffio[a], Huihao Zhang[o], Aaron Brown[a], Jason Fucik[b], Aidan Gibbs[d], Rose Gibson[d], Sam Halverson[c], Christopher Johnson[d], Sonia Karkar[g], Takayuki Kotani[h,p], Evan Kress[d], Stephanie Leifer[q], Kenneth Magnone[d], Jerome Maire[a], Rishi Pahuja[b], Michael Porter[b], Mitsuko Roberts[b], Ben Sappey[a], Jim Thorne[g], Eric Wang[d], Étienne Artigau[j], Geoffrey A. Blake[b], Gabriela Canalizo[r], Guo Chen[s], Greg Doppmann[g], René Doyon[j], Courtney Dressing[t], Min Fang[s,u], Thomas Greene[v], Greg Herczeg[w], Lynne Hillenbrand[b], Andrew Howard[b], Stephen Kane[r], Tiffany Kataria[c], Eliza Kempton[x], Heather Knutson[b], David Lafrenière[j], Chao Liu[y], Stanimir Metchev[z], Max Millar-Blanchaer[aa], Norio Narita[h,i], Gajendra Pandey[bb], S.P. Rajaguru[bb], Paul Robertson[cc], Colette Salyk[dd], Bun'ei Sato[ee], Evertt Schlawin[ff], Sujan Sengupta[bb], Thirupathi Sivarani[bb], Warren Skidmore[f], Gautam Vasisht[c], Chikako Yasui[f], Hui Zhang[gg]

[a]Department of Astronomy & Astrophysics, University of California, San Diego, 9500 Gilman Drive, La Jolla, CA, 92093, USA; [b]California Institute of Technology, Pasadena, CA, 91125, USA; [c]Jet Propulsion Laboratory, California Institute of Technology, Pasadena, CA 91109, USA; [d]Department of Physics & Astronomy, University of California, Los Angeles, Los Angeles, CA 90095, USA; [e]NASA Exoplanet Science Institute, Infrared Processing and Analysis Center, Jet Propulsion Laboratory, California Institute of Technology, Pasadena, CA 91125, USA; [f]Thirty Meter Telescope International Observatory, Pasadena, CA 91124, USA; [g]W. M. Keck Observatory, Kamuela, HI, 96743, USA; [h]The University of Tokyo, Tokyo, Japan; [i]Astrobiology Center, National Institutes of Natural Sciences, Tokyo, Japan; [j]Department of Physics and Institute for Research on Exoplanets, Université de Montréal, Montreal, QC, Canada; [k]Department of Astronomy, University of Wisconsin, Madison, Madison, WI, 53706, USA; [l]Miyagi University of Education, Sendai, Miyagi, Japan; [m]Department of Physics and Astronomy, George Mason University, Fairfax, VA, 22030, USA; [n]Center for Interdisciplinary Exploration and Research in Astrophysics (CIERA) and Department of Physics and Astronomy, Northwestern University, Evanston, IL 60208, USA; [o]Department of Astronomy, The Ohio State University, Columbus, OH 43210, USA; [p]Subaru Telescope, Hilo, HI 96720, USA; [q]Aerospace Corporation, El Segundo, CA 90245; [r]University of California, Riverside, Riverside, CA 92521, USA; [s]Purple Mountain Observatory, Chinese Academy of Science, Nanjing 210023, China; [t]Department of Astronomy, University of California, Berkeley, Berkeley, CA, 94720; [u]University of Science and Technology of China, Hefei 230026, China; [v]Space Science and Astrobiology Division, NASA's Ames Research Center, Moffet Field, CA 94035, USA; [w]Peking University, Beijing 100871, China; [x]Department of Astronomy, University of Maryland, College Park, MD, 20742, USA; [y]National Astronomical Observatories of China, Chinese Academy of Sciences, Beijing, 100012, China; [z]Department of Physics and Astronomy, Western University, London, ON, Canada; [aa]Department of Physics, University of California, Santa Barbara, Santa

---

[*] Further author information: (Send correspondence to Quinn Konopacky)
Quinn Konopacky: qkonopacky@ucsd.edu

Barbara, CA 93106, USA; [bb]Indian Institute of Astrophysics, Bangalore, 560034, India; [cc]Department of Physics & Astronomy, University of California Irvine, Irvine, CA 92697, USA; [dd]Department of Physics and Astronomy, Vassar College, Poughkeepsie, NY 12604, USA; [ee]Department of Earth and Planetary Sciences, Tokyo Institute of Technology, Tokyo 152-8550, Japan; [ff]Steward Observatory, University of Arizona, Tucson, AZ, 85721, USA; [gg]Shanghai Astronomical Observatory, Chinese Academy of Sciences, Shanghai 200030, China

**ABSTRACT**

HISPEC is a new, high-resolution near-infrared spectrograph being designed for the W.M. Keck II telescope. By offering single-shot, R 100,000 spectroscopy between 0.98 – 2.5 μm, HISPEC will enable spectroscopy of transiting and non-transiting exoplanets in close orbits, direct high-contrast detection and spectroscopy of spatially separated substellar companions, and exoplanet dynamical mass and orbit measurements using precision radial velocity monitoring calibrated with a suite of state-of-the-art absolute and relative wavelength references. MODHIS is the counterpart to HISPEC for the Thirty Meter Telescope and is being developed in parallel with similar scientific goals. In this proceeding, we provide a brief overview of the current design of both instruments, and the requirements for the two spectrographs as guided by the scientific goals for each. We then outline the current science case for HISPEC and MODHIS, with focuses on the science enabled for exoplanet discovery and characterization. We also provide updated sensitivity curves for both instruments, in terms of both signal-to-noise ratio and predicted radial velocity precision.

## 1. INTRODUCTION

Thirty years after the discovery of the first planet orbiting a star other than the Sun, it is now clear that extrasolar planets are ubiquitous in the galaxy. Though our understanding of exoplanet demographics has dramatically expanded in recent years, many fundamental questions remain about their elemental compositions as well as their formation and evolution pathways.[1] In addition, we also have a highly incomplete understanding of the chemical and physical processes shaping exoplanet atmospheres today, including the role of disequilibrium chemistry, the formation of clouds and hazes, as well as the dynamical process under extreme conditions.[2,3,4]

The next decade presents a truly unique opportunity to address many of these questions. For the first time, we have the observational techniques, the theoretical models, and a sufficient number of known exoplanets orbiting nearby stars to spectroscopically characterize a wide diversity of planets – from hot giant planets to temperate Earth-size planets. However, these observations are extremely challenging. Teasing out the small exoplanet signal from the stellar photon shot noise and calibrating systematics require both high sensitivity and spectro-photometric precision.

Recent exoplanet studies have largely focused on photometric or low-resolution spectroscopic measurements. This research (e.g., *HST* and *Spitzer* studies of transiting planets, and adaptive optics photometry and spectroscopy of directly imaged planets) has yielded critical new insights into the bulk properties of planet atmospheres at both close[5] and wide separations.[6] Indeed, *JWST* has inherited this mantle, offering a new observing regime for a wide range of exoplanets.[7,8] However, *JWST*'s maximum R~3,000 spectral resolution will not be able to explore the full range of exoplanet science that is critical to advancing the field. This is because, at low spectral resolution (R≲5,000), overlapping spectral bands can lead to degeneracies, minor chemical species important to understand the chemical processes often remain masked, and clouds substantially mute broad spectral signatures.[9,10,11] Furthermore, at low to modest spectral resolution the spectral line position and line shape information necessary to detect and constrain exoplanetary winds is smeared out.

High-resolution spectroscopy (R~100,000) has recently emerged as a powerful tool for exoplanet studies,[12,13,14,15,16,17,18,19,20,21] providing highly complementary information to low- and moderate-resolution observations from space.[22,23] High spectral resolution offers the opportunity to distinguish the planet signal from stellar and telluric contamination via the wealth of information contained in the *resolved and Doppler-shifted* NIR atomic and molecular features in a planet's atmosphere. With an appropriate design, a NIR high-resolution spectrograph on a large, ground-based telescope enables **all three major methods of exoplanet detection and characterization (transit spectroscopy, direct imaging, and radial velocity)**. This is the fundamental driver behind the design of the High-resolution Infrared Spectrograph for Exoplanet Characterization (HISPEC) and the Multi-Objective Diffraction-limited High-resolution Infrared Spectrograph (MODHIS). These two instruments, the first for the W.M. Keck Observatory and the second for the Thirty Meter Telescope, will offer a paradigm shift in observational studies of exoplanets. Additionally, both HISPEC and MODHIS will enable general astrophysics investigations complementary to exploring the exoplanet frontier. Additional science cases include studies of stars in the Galactic Center, probing the dynamics of stars in other crowded regions such as star-forming regions and clusters, measuring abundances of stars in the Galactic bulge, halo, and nearby dwarf galaxies, unique studies in the Solar System, and potentially spectropolarimetry.

## 2. INSTRUMENT OVERVIEW

Though designed for different telescope facilities, HISPEC and MODHIS share many similarities that make them compatible with a coeval design process. In this section, we briefly describe the current design of each instrument. For additional details, we refer the reader to Mawet et al. [24].

### 2.1 HISPEC

HISPEC is under development for the W.M. Keck II 10-meter telescope. The instrument is subdivided into several major subsystems. The instrument is fed by the Keck II facility AO system, which routes light into the Front-End Instrument (FEI), which is responsible for acquisitions, fiber injection, and precision guiding. The FEI then interfaces with the Fiber Subsystem (FIB), which also accepts light from the Calibration Subsystem (CAL). The FIB is responsible for delivering all light to the red spectrograph and the blue spectrograph. The system is controlled using the Instrument Control Software and spectra are processed using the Data Reduction Pipeline (DRP). Figure 1 provides a basic block diagram overview of each of the main subsystems of HISPEC.

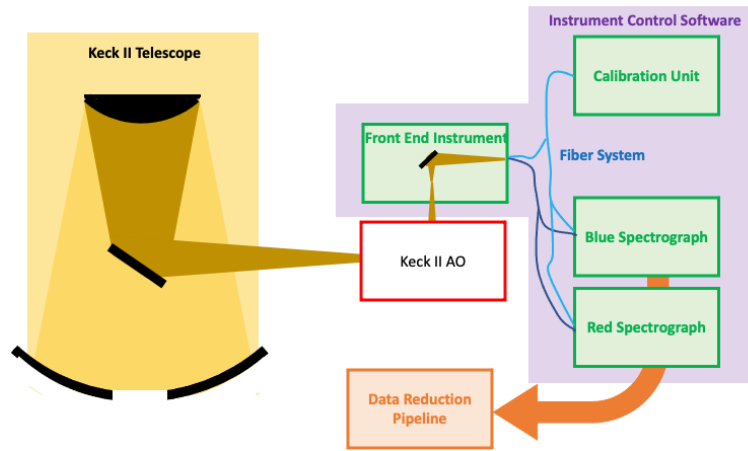

Figure 1. Basic block diagram of HISPEC. Light from the telescope and AO system is fed to the Front-End Instrument (FEI), and then routes to the spectrographs (BSPEC and RSPEC) via the Fiber System (FIB). The whole system is controlled by the Instrument Control Software (SW). Data from the spectrographs is processed via the Data Reduction Pipeline (DRP).

The front-end instrument (FEI) for HISPEC will be the primary instrument subsystem with which a science user will interact. Past the AO system components, which are operated by observatory staff, all moving components of the instrument are housed in the FEI, FIB, or CAL subsystems. Of primary importance is the acquisition and pupil imaging camera (ATC) that will provide real-time information about the positioning of the science field-of-view relative to the fiber entrances. ATC will consist of a small Dewar housing a H2RG with a goal 6"x6" (9.4 mas/pixel) field of regard on the science target. The science user will be able to select at which wavelength they wish to track the science field-of-view, using movable mechanisms including stages containing different dichroics (ATC pick-off) and a filter wheel in the ATC Dewar path. Steering of the field-of-view with respect to the fiber input will be controlled by a field steering mirror. Software will provide continuous tracking and correcting commands of the centroid of the guide star. Manual adjustments will also be possible. A mask selector will provide the user with options for additional pupil masks, such as a circular pupil stop, vortex fiber nuller[25,26], or a gray scale apodizer. Phase Induced Amplitude Apodization (PIAA) optics will also be selectable depending on the type of observation and will move in and out of the beam on stages. The corner cube is used for retroreflection, sending light from the fibers back to the ATC field-of-view to provide the exact location of the fiber tip.[27] This will likely be primarily used during daytime calibrations, and the proper pixel location for the science image will be recorded and used in software. The ADC movement will follow the motion of the telescope for compensation as a function of zenith angle.

The fiber delivery system (FIB) will perform several functions. It will guide light from the FEI to the spectrographs housed in the basement, some distance from the telescope. It will also provide the means to feed light from the calibration subsystem to the HISPEC spectrographs, including simultaneous wavelength solution tracking with a laser frequency comb (LFC) or a Fabry-Perot Etalon, absolute wavelength calibration light, and detector flatfield light. Additionally, it will provide the means for retroreflection into the FEI to pinpoint the location of the science fiber on the ATC. It will accomplish these tasks using a series of robotically controlled switches.

The spectrograph dewars (BSPEC and RSPEC) will be housed in the basement of Keck, far from the telescope. This is to maximize their stability. There are no choices to be made by a user related to the spectrographs other than exposure time and readout setting. All moving parts are upstream of the spectrographs. User choices about tracking and guiding may result in not all wavelength regimes being available for spectroscopy.

The calibration subsystem (CAL) is a critical portion of the instrument. The scientific goals of high precision radial velocity for detection of exoplanets requires a detailed and critical assessment of calibration needs (§3). This system will consist of multiple wavelength and line spread function calibration sources, including a laser frequency comb[28], a Fabry-Perot etalon, and lamps with known molecular or atomic species. It will also contain the detector flatfield lamps. Users will have the option to select which wavelength reference source they would like to inject into the spectrograph simultaneously with their science data.

Like the other PRV instruments, HISPEC will require a sophisticated data reduction pipeline. The DRP will fit in to the Keck Data Services Initiative Framework and have a Quick-look version for on-the-fly analysis at the telescope. For the purposes of HISPEC, we have identified Level 0, 1, and 2 data products that may be processed and/or output by the DRP. Observer sequences produce Level 0 data products, which comprise raw echellogram images as well as associated metadata from the telescope, AO system, and FEI. Level 1 data products are the those that have undergone some amount of data processing, and hence should include 1-dimensional spectra, calibrated for wavelength, instrumental response, and flux. Level 2 data products are fully processed, science-ready products that are dependent on the specific science goals. Outputs may include precise RVs, starlight-suppressed directly imaged planet spectra, derived planet spectra from transit observations, or fully flux-calibrated, telluric corrected spectra across multiple selectable spectral orders.  The HISPEC DRP team is currently evaluating the best means to provide higher level data products to the wider community.

## 2.2 MODHIS

MODHIS is under development for the Thirty Meter Telescope.  The MODHIS instrument will be mounted to the TMT facility AO system NFIRAOS and can be divided into several major subsystems. The interface to NFIRAOS is provided by the Cable wrap, Support structure, Rotator, and On-Instrument Wavefront Sensor (OIWFS) unit, which then interfaces to the Front-End Instrument (FEI), which itself consists of the Calibration Unit (CAL), and the Fiber Injection Unit (FIU). The FEI/FIU interfaces to the fiber delivery subsystem (FIB) which transfers science (and calibration) light to the red spectrograph (RSPEC) and blue spectrograph (BSPEC) subsytems, located in the TMT basement. Ultra-stable calibration signals are provided by a laser frequency comb subsystem. The system is controlled using the Instrument Control Software (ICS) and spectra are processed using the Data Reduction Pipeline (DRP). Figure 2 provides a block diagram overview of each of the main subsystems of MODHIS.

The Narrow-Field InfraRed Adaptive Optics System (NFIRAOS) is the first-light facility AO system for TMT. It will deliver a large, well-corrected, uniform field with high sky coverage and minimal extra thermal background to multiple science instruments, including MODHIS. More specifically, NFIRAOS is a Multi-Conjugate AO system with 2 deformable mirrors and 6 laser guide star wavefront sensors that pass a two-arcminute field with an expected average wavefront error of 203 nm (53% Strehl ratio in H-band and 71% Strehl ratio in K-band) over the central 34"x34". Simulations show that NFIRAOS with an OIWFS can achieve 50% sky coverage, even at the North Galactic Pole. NFIRAOS resides in an enclosure that is cooled to -30 C to ensure that the thermal background from NFIRAOS does not exceed 15% of that of the sky plus telescope at any wavelength. The NFIRAOS throughput will be greater than 85% at wavelengths greater than 1 micron. The design of NFIRAOS is well-documented and can be found in several references.[29]

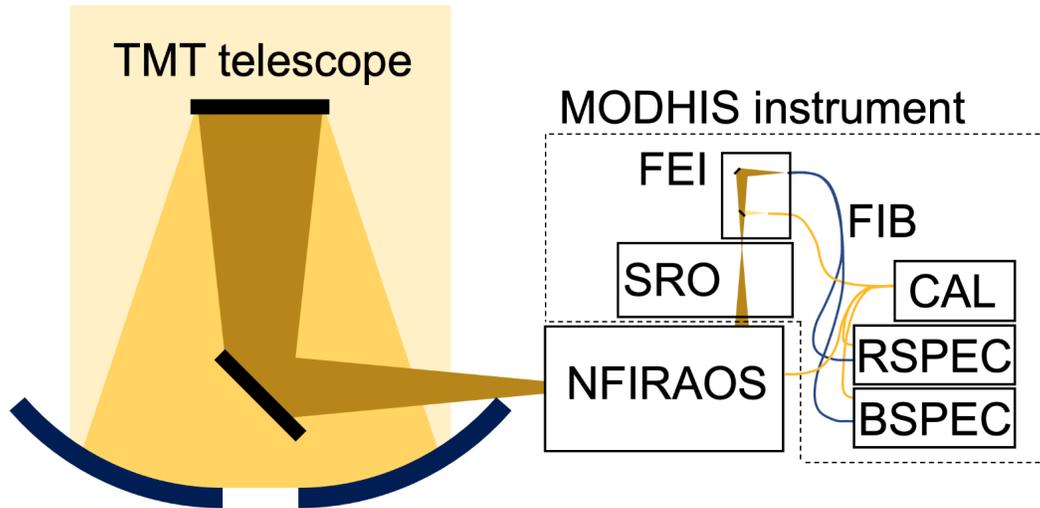

Figure 2. Block diagram of the MODHIS optical path. The light from NFIRAOS passes through the key optical sub-systems of MODHIS: the SRO, CAL, FEI, FIB, and RSPEC and BSPEC.

MODHIS includes a cold enclosure (-30º C) that houses three low-order On-Instrument WFS's (OIWFS) that are be used by the NFIRAOS RTC to sense tip-tilt and focus. Each will sample stars as faint as J < 21-22 over a 2 arcminute Field of Regard. The WFSs are each deployable over roughly half of their total field of regard in order to maximize sky coverage and achieve the optimum AO correction. The current design is baselined from the IRIS OIWFS design, which is described in Atwood et al. [30]. MODHIS must rotate about the optical axis in order to track field rotation, and to select position angles. The Structure, Rotator, OIWFS (SRO) therefore includes a large rotator bearing that rotates both the OIWFS enclosure and front-end instrument, which ports the light from the telescope via fibers to the spectrographs in the basement. MODHIS will be on the top port of NFIRAOS, and the rotation requires a cable wrap to keep the many cables and hoses from tangling or getting damaged during observations. The design of the rotator and wrap is still preliminary, with the goal of allowing continuous sky tracking for many hours before requiring a de-wrap.

Past NFIRAOS and the SRO/OIWFS components, which are likely to be operated by observatory staff, all moving components of the instrument are housed in the FEI or CAL subsystems. Of primary importance is the acquisition and pupil imaging camera (APIC) that will provide real-time information about the positioning of the science field-of-view relative to the fiber entrances. APIC will consist of a small dewar housing a H2RG with a roughly 6"x6" (2.75 mas/pixel) field of regard on the science target. The science user will be able to select at which wavelength they wish to track the science field-of-view, using movable mechanisms including a wheel containing different beam splitters and potentially a filter wheel inside of the APIC dewar. Steering of the field of view with respect to the fiber input will be controlled by the tip/tilt mirror. Software will provide continuous tracking and correcting commands of the centroid of the guide star. Manual adjustments will also be possible. A mask selector will provide the user with options for additional pupil mask, such as a circular pupil stop (likely the default selection), vortex fiber nuller, or a gray scale apodizer. PIAA optics will also be selectable depending on the type of observation and will move in and out of the beam on stages. The corner cube is used for retroreflection, sending light through the fibers back to the APIC field-of-view to provide the exact location of the fiber entrance. This will likely be primarily used during daytime calibrations, and the proper pixel location for the science image will be recorded and used in software. The ADC movement will follow the motion of the telescope for compensation as a function of zenith angle

The calibration system is envisioned to be very similar to HISPEC, with multiple wavelength and line spread function calibration sources, including a laser frequency comb, a Fabry-Perot etalon, gas cell lamps with known molecular species, and detector flatfield lamps. The primary difference between HISPEC and MODHIS will be that MODHIS may also offer polarimetric observations. If that mode is included, additional polarization calibration infrastructure will be included, both in the FEI and potentially in the NFIRAOS Science Calibration Unit (NSCU). The FIB system is also similar to what is described for HISPEC, wherein it will be responsible for routing both science light and calibration light to the RSPEC and BSPEC subsystems. The location of RSPEC and BSPEC is still under study. Potential locations include the Nasmyth platform and a room under the telescope known as the Wedge Room. A trade study will determine which location offers more stability for the PRV goals of MODHIS. As with all first light instruments for TMT, MODHIS will have a DRP to process the datasets. The DRP has not yet been designed, but will follow the protocol and structure of DRPs defined by the observatory.[31]

### 2.3 Instrument Requirements

The designs of both HISPEC and MODHIS flow from the science cases described in Section 3. The scientific goals of the instruments are ambitious and multi-faceted. In particular, we are designing the instrument to optimally achieve multiple scientific goals in the realm of exoplanets, including PRV, transit spectroscopy, and direct spectroscopy. It is possible to design instruments that can achieve all of these aims, as we will demonstrate. Table 1 provides an overview of some of the high-level requirements for both HISPEC and MODHIS.

Table 1. A selection of the high-level requirements based on science cases for both HISPEC and MODHIS.

| | HISPEC (Keck) Requirement(s) | MODHIS (TMT) Requirement(s) | Notes/Rationale |
|---|---|---|---|
| Wavelength Coverage | - The system shall provide spectroscopic wavelength coverage between 0.98 – 2.46 μm. | -Wavelength coverage of MODHIS shall be 0.98-2.46 μm. | Precision RV requires 0.98 – 2.46 μm to achieve maximum Doppler content and sample regions with minimal telluric contamination<br><br>Exoplanet transmission spectroscopy requires 0.98 – 2.46 μm to maximally sample molecular features<br><br>The blue end of the spectrum between 0.98 - 1.0 μm has several unusual lines that can be used for precision abundance work |
| Image Quality | -The acquisition camera shall provide a diffraction-limited core at all wavelengths.<br><br>-The spectrograph internal image quality shall not impact the theoretical resolving power by more than 10% across the bandpass. | -MODHIS high order wavefront error at fiber injection shall be less than 40 nm RMS after correction by NFIRAOS DM0.<br><br>-The spectrograph internal image quality shall not degrade the theoretical resolving power by more than 10% across the bandpass. | Maximized coupling efficiency is critical for faint sources, such as off-axis directly imaged planets. The acquisition camera provides the check on alignment of the fiber core with the PSF core. |
| Spatial Sampling | -The acquisition camera shall achieve a plate scale that Nyquist samples the diffraction limit at 1.02 μm).<br><br>-The plate scales at the fiber focal planes shall be adjusted to optimally couple light into the fibers at all wavebands.<br><br>-The average sampling of the spectrographs shall be >3 pixels. | -The acquisition/tracking camera sampling shall be sufficient to Nyquist sample the diffraction limit at 0.82 μm.<br><br>-MODHIS shall provide pixel sampling across spectral orders of at least 3.0 pixels per FWHM of the spectral line with, over 80% of total spectrum captured by each SPEC channel. | Precise alignment of the input PSF and the fiber core is required to maximize the SNR of all targets.<br><br>Decades of experience with precision RV measurements in the optical have shown super-Nyquist wavelength sampling maximizes RV return |
| Field of Regard | -The tracking camera shall provide a circular field-of-view of at least 5" diameter<br><br>-The object fiber shall be capable of moving to any location in the tracking camera | -The field of view of the MODHIS tracking camera shall be no less than 6 x 6 arcsec square.<br><br>-Each spectrograph science fiber shall be capable of relative positioning at location in the | Ease of target acquisition is provided by a modest tracking camera field-of-view for all science cases<br><br>High contrast targets require guiding on the primary star that is offset from the planet by <2". |
| | field-of-view. | tracking camera field-of-view.<br>-The SRO shall provide a field rotator that rotates the science field 540 degrees full range about the telescope optical axis. | Galactic Center observations at K band will require guiding on off-axis targets that are brighter at H band |
| Spectral Resolving Power | -The spectrographs will achieve an average resolution of R=100,000 across the 0.98-2.46 μm range. | -MODHIS shall provide high-resolution for each of the red and blue channels with R ≥ 100,000 (averaged across each channel's waveband separately), assuming a diffraction-limited fiber input. | The number of transiting planet targets observable increases with increased spectral resolution |
| System Efficiency | -The spectrograph shall provide SNR>30 per resolution element at YJHK on a magnitude 15 star in 4 hours of observing. | -The spectrograph shall provide SNR>17 per resolution element at YJHK on a magnitude 15 star in 4 hours of observing. | Typical directly imaged planets are ~15-17 mags at K band, and SNR~30 provides needed sensitivity to atmospheric and kinematic parameters<br><br>Precise RVs require high SNR observations in reasonably short exposure times<br><br>Transit spectroscopy requires high SNR observations during a transit timescale |
| Multiplexing | -The fiber injection unit shall provide at least three channels: 1 object, 1 speckle, and 1 background.<br>-The instrument shall accommodate a calibration fiber that feeds the spectrograph. simultaneously with object and background measurements.<br>-All spectral orders shall be wide enough to accommodate non-overlapping traces from *at least* four fibers (1 object, 1 background, 1 speckle, 1 calibration). | -MODHIS shall provide a high-dispersion, high-contrast (HDHC) observing mode wherein science target, stellar speckle light, sky/background, and reference frequency light will be concurrently observed.<br>-MODHIS shall provide a precision radial velocity (PRV) observing mode wherein science target, sky/background, and reference frequency light will be concurrently observed.<br>-The layout of spectral traces shall maintain an average cross-coupling intensity at any point in a trace (averaged as if originating from the entire detector) of no more than 0.1%. | Two simultaneous fiber channels can sample a directly imaged planet and the speckle field. The background fiber will sample the background noise from sources such as thermal/sky and dark current.<br><br>A calibration channel offers required wavelength stability for precision RV and for LSF monitoring for Doppler imaging |
| Radial Velocity Precision | -The internal velocity precision of the instrument shall be <30 cm/s and shall maintain this stability over a timescale of years.<br>-The absolute spectral stability on the detector shall be 0.01 pixels and shall be maintained over a timescale of nights. | -MODHIS systematic errors shall not preclude the calibrated measurement of stellar radial velocities with ≤ 30 cm/s accuracy, ignoring the contribution from photon noise. | Required to achieve on-sky RV precision goal of ~1 m/s<br><br>A pixel stability of 0.01 avoids intra-pixel flat-fielding effects that impact RV precision. |
| Simultaneous Spectral Coverage | -The blue spectrograph shall provide simultaneous coverage across the YJ band.<br>-The red spectrograph shall provide simultaneous coverage | -MODHIS shall provide a blue spectroscopy channel with wavelength coverage 0.98 μm - 1.33 μm (YH bands).<br>-MODHIS shall provide a red | Observing efficiency is roughly linearly proportional to the simultaneous wavelength coverage<br><br>Simultaneous observations in J and K bands will allow detailed studies of |

| | across the HK band. | spectroscopy channel with wavelength coverage 1.49 μm - 2.46 μm (H+K bands). | variability in brown dwarfs and giant planet |
|---|---|---|---|
| Data Quality and Data Reduction System | -The data reduction system shall produce rectified and wavelength-calibrated 1-D spectra from all spectral orders<br><br>-Reduced data sets shall achieve relative photometry between wavelength channels with systematic error of ≤1% | -The MODHIS DRP shall produce rectified and wavelength- calibrated 1-D spectra from all spectral orders<br><br>-Reduced data sets shall achieve relative photometry between wavelength channels with systematic error of ≤1%. | Level 1 data products offer uniformity in reduction and processing<br><br>Transit spectroscopy requires high precision measurements of line depths that are calibrated across orders<br><br>Transit spectroscopy requires minimized flat-fielding errors<br><br>Precision RV and stellar abundance measurements require high precision blaze correction |
| Polarimetric Capabilities | The spectrograph design shall not preclude future polarimetric mode observations. | MODHIS should provide spectropolarimetry with 0.1% (goal) polarimetric precision. | If offered, MODHIS will be the only first light ELT instrument with spectropolarimetric capabilities. |

## 2.4 Predicted Sensitivity

Based on the requirements given in the previous section, the instrument design has advanced past the Preliminary Design phase for HISPEC and the Conceptual Design phase for MODHIS. This provides a good estimate of the expected instrument throughput for both instruments. We also have precise understanding of the sky and telluric backgrounds (assuming Maunakea as a baseline for both telescopes), telescope and AO system emissivity, and AO system wavefront error. With these pieces of information and the current instrument designs, we have generated a robust end-to-end simulator that can predict the expected signal-to-noise and RV precision for a given target, utilizing spectral models of stars and/or planets. A more detailed description of some of these simulation tools can be found in Gibbs et al. [32].

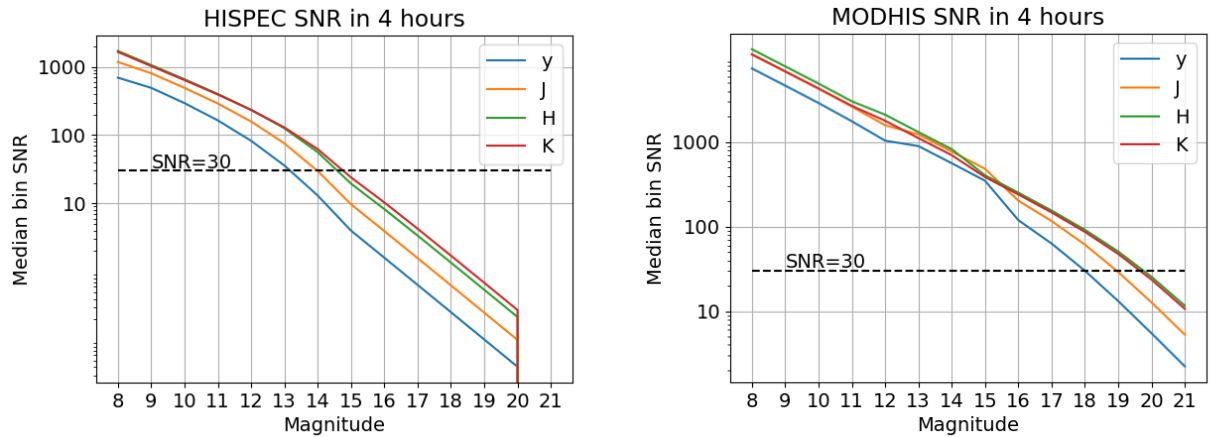

Figure 3. Predicted sensitivity plots for point sources observed with HISPEC (left) and MODHIS (right), presented as a function of waveband. The plots show the median SNR across all spectral elements for each waveband. Because sensitivity the requirements in Table 1 are tied to an SNR~30, we also denote the location at which the curves cross that line. For MODHIS, we are exceeding the spec of SNR~30 at 17th magnitude for all wavebands. For HISPEC, the SNR~30 is met for the red spectrograph (H and K), but not currently for Y and J. The instrument team is currently assessing what can be done to increase sensitivity in the blue spectrograph for HISPEC to meet this requirement.

Using this simulation tool, we are able to compute the expected point source sensitivity as a function of magnitude. We show these sensitivities for both HISPEC and MODHIS in Figure 3. Note that this represents the median sensitivity per bin across the full wavelength range of the instruments. Based on our current best estimates, we will achieve the required sensitivity as listed in Table 1 with MODHIS, but may not reach the desired performance at short wavelengths with HISPEC. Investigations into whether this can be improved are ongoing. Still, both instruments should be quite sensitive for high resolution spectrographs.

## 3. SCIENCE CASES

The science cases for HISPEC and MODHIS have been derived by the science teams in conjunction with the instrument technical teams and observatory staff. While the focus of both instruments has been on advancing exoplanetary science, the scientific utility of both will expand well-beyond exoplanets. Here, we describe some of our current science cases and how both instruments will offer a new suite of capabilities.

### 3.1 The Atmospheres of Close-in Exoplanets

Heated by their host stars, close-in exoplanets (orbital separations of <0.01 AU) have prominent observable atmospheric features, which make them ideal targets for characterization. Indeed, some of the most detailed data we have on planetary atmospheres outside the Solar System comes from hot Jupiters.[33] This will continue to be the case for at least the next decade.

The general goal of these types of atmospheric observations is to measure both the bulk abundances of the planetary atmospheres, as well as specific elemental abundance ratios (usually C/H, O/H, and N/H). These two types of measurements allow us to test theories of planet formation via trends in bulk metallicity[34] and to directly probe the planet formation pathways via elemental abundance measurements.[35] Atmospheric observations of close-in exoplanets can also identify potentially habitable, warm and temperate, atmospheres on super-Earths orbiting lower-mass M-dwarfs.[36]

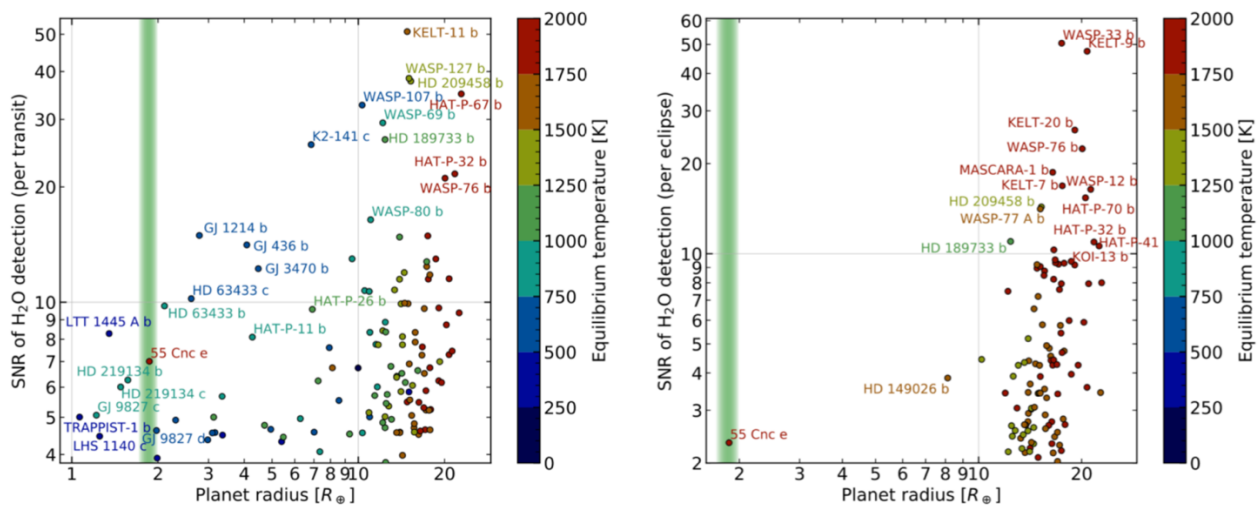

Figure 4. Potential targets for transit observations (left) and secondary eclipse observations (right) with HISPEC. All labeled planets represent the most favorable targets among all known transiting planets with declinations observable from Maunakea (circles). The signal-to-noise ratio of water detections in a hydrogen-dominated atmosphere is indicated versus the planet radius, with the color indicating the equilibrium temperature. SNRs are calibrated using true injection-recovery tests using cross-correlation analyses on sequences of realistically simulated Keck/HISPEC exposures that included tellurics and instrumental noise. Planets below 5 $R_\oplus$ must be observed 3 to 8 times over multiple observing cycles to build up SNR to probe for high mean molecular mass (e.g., GJ 9827c, 55 Cnc e) and highly hazy/cloudy atmospheres (e.g., GJ 1214b). The green vertical line indicates the position of the radius valley in the small planet occurrence rates believed to separate rocky planets and sub-Neptunes.[38]

One method to measure the atmospheres of close-in exoplanets is using transit and eclipse spectroscopy of transiting planets. HISPEC and MODHIS transit observations of close-in exoplanets provide a unique tool to explore the compositional diversity of planets and their atmospheres, by determining the bulk elemental compositions of giant planet envelopes. Transit spectroscopy also explores the nature of clouds and hazes near the terminator on these planets, and can study the atmospheres of sub-Neptunes and terrestrial exoplanets — plausibly even with temperate conditions.

HISPEC and MODHIS will also be able to conduct high-resolution spectroscopic observations of close-in exoplanets to measure atmospheric abundances and temperature structure and can do so for both transiting and non-transiting exoplanets. Currently, exoplanet atmospheric characterization using these types of high-resolution spectroscopy observations cross-correlate against the entire planetary spectrum,[12] which yield good overall molecular abundances. HISPEC will particularly excel for cloudy and hazy atmospheres (e.g., GJ 1214b) because the narrow cores of molecular lines observable at high resolution (R>80,000) form above the clouds,[10,11] unlike the broad molecular bands observable at low and medium resolution with *JWST* which are muted by the presence of clouds.[9] In addition, the narrow planetary molecular lines shift in velocity during the transit, enabling an unambiguous distinction from any $H_2O$ or CO stellar or telluric lines.[37] Figure 4 demonstrates potential targets and the expected SNR that can be achieved with HISPEC.

Excitingly, the large aperture of TMT and the high sensitivity of MODHIS mean that we will be able to conduct these same observations on a line-by-line basis, thereby constructing a real high-resolution spectrum of the planetary atmosphere. We can then use the relative strengths of these individual lines in a molecular band to directly constrain the temperature structure of a planets' atmosphere, along with measuring specific elemental abundances. By the time MODHIS begins its observations, the Transiting Exoplanet Survey Satellite (TESS) and PLATO will have already explored the entire sky to identify the most favorable exoplanet targets, and many of these planets will have very precise, but low resolution (100<R<1000), atmospheric spectra from JWST. This will offer an ideal opportunity for higher resolution studies with MODHIS. A similar analysis for MODHIS as performed for HISPEC that is demonstrated in Figure 4 shows an expectation of hundreds of targets could be observed with SNR > 5.

### 3.2 Chemodynamics of Resolved Substellar Companions

The question "How do giant planets form?" is best answered by characterizing them near the epoch of their formation. This can be achieved on Solar System scales (<40 AU) with HISPEC and MODHIS, which combine high angular resolution, high contrast, and high-resolution spectroscopy,[15,17,39,40] a technique also known as High Dispersion Coronagraphy (HDC).[41,42] The AO system separates the light from the star and the companion, whose signals are then individually fed to the spectrograph via SMFs for further spectral differentiation. Furthermore, using HDC as a pathway to habitable worlds with the ELTs is one of the four key capabilities laid out by Astro2020 (see their §1.1.1).[43] The techniques developed by HISPEC for HDC science cases will shape Earth-finding instruments like MODHIS on TMT.

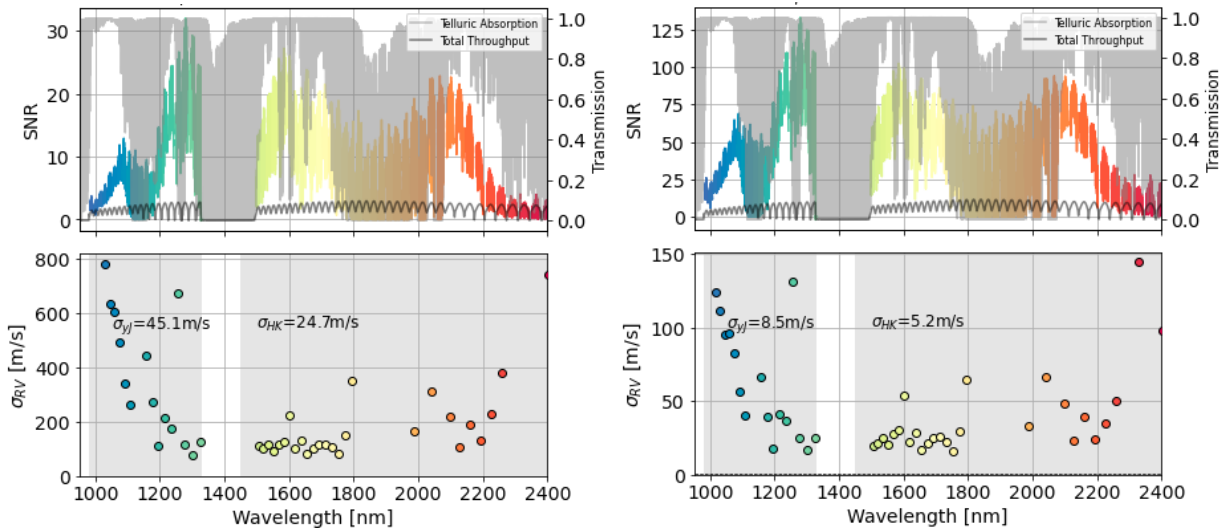

Figure 5. Sensitivity of HISPEC and MODHIS for direct spectroscopy of exoplanet HR 8799c, calculated using our end-to-end instrument simulator. The top panels show the expected SNR across the full spectral range covered by both instruments, in addition to demonstrating the assumed telluric signal and throughput. The bottom panels show the predicted RV precision that can be achieved at specific wavelengths. HR 8799c has a K band magnitude of ~16 and is significantly impacted by speckles. The SNR that can be achieved with HISPEC and MODHIS demonstrates the power of using HDC to reject speckle noise. The predicted RVs precisions are impressive, and for MODHIS are small enough for the detection of Solar System-like exomoons (§3.3.3).

HISPEC and MODHIS will be able to survey directly imaged giant planets beyond a few AU, enabling the spectroscopic characterization of potentially 100s of giant planets. This range of separations includes the peak in giant exoplanet occurrence rates at ~2.5 AU as measured by RV surveys,[44,45] allowing us to study the most typical giant planets and place our own Jupiter into context. By studying the shapes of spectral lines at high spectral resolution, HISPEC and

MODHIS will be able to measure the compositions, spins, and orbits of typical giant planets at a population level. Figure 5 demonstrates the sensitivity of both instruments using the well-studied directly imaged planet, HR 8799c. Models of core accretion and late stage planetesimal accretion predict gradients in the metallicities and elemental abundance ratios of giant planets as a function of mass and semi-major axis.[46] Current leading theories that magnetic breaking in the protoplanetary disk sets the final spin rates of giant planets predicts gradients in spin speeds as a function of mass.[47] Measuring the eccentricity distribution of typical giant planets will inform us whether giant planets create dynamically quiescent planetary systems like our own. HISPEC and MODHIS will empirically find the trends that exist in these planet properties that will serve as some of the only constraints to highly complex planet formation models.

Additionally, HISPEC and MODHIS will be able to characterize planets at various key stages in their formation and evolution: 1) newborn (1 Myr) protoplanets residing in nearby (~140 pc) star-forming regions to witness planet formation in action; 2) adolescent (~30 Myr) planets that recently finished forming and are still radiating away the signatures of their formation process; 3) mature (> 1 Gyr) planets. HISPEC and MODHIS will be able to study the time evolution of planetary characteristics. Planetary orbits evolve through gravitational interactions and the amount orbits change informs us how dynamically hot planetary systems are. Atmospheric composition changes due to accretion of solids and gases from the circumstellar disk. If magnetic breaking with the protoplanetary disk is indeed responsible for setting planetary spin rates, protoplanets likely have a different spin distribution than older planets for which angular momentum is conserved.

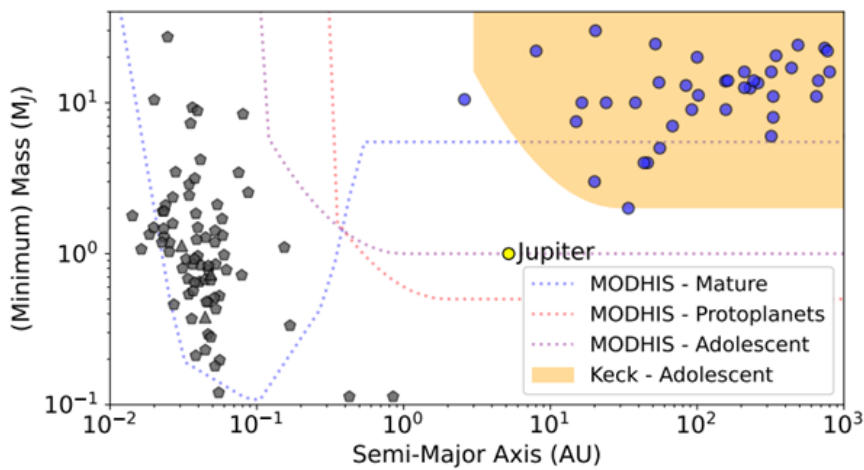

Figure 6. Sensitivity of HISPEC and MODHIS for direct spectroscopy of exoplanets for three different classes of planets: protoplanets (1 Myr old) residing in the closest (140 pc) star-forming regions, adolescent (40 Myr) planets residing in nearby (30 pc) young moving groups, and mature (1 Gyr) planets residing in the Solar neighborhood (10 pc). For protoplanets and adolescent planets, the planets are dominated by the glow from their heat of formation. For mature planets close-in (< 1 au), thermal heating and reflected light become dominant contributors, allowing for increased sensitivity close-in. HISPEC will excel at detecting young, more widely separated planets, while MODHIS will probe true Jupiter analogs.

While the majority of these planets are yet to be discovered, HISPEC and MODHIS is capable of either discovering them or following up indirectly discovered planets. The Gaia astrometric mission will uncover thousands of new giant exoplanets regardless of their age.[48] Radial velocity has discovered many mature planets beyond 1 au,[45] and precise infrared spectrographs like HISPEC and MODHIS itself will find adolescent giant planets. MODHIS will focus on follow-up characterization of these indirectly discovered planets. Both instruments will also have a discovery mode: the vortex fiber nuller (VFN).[49,50] VFN combines nulling interferometry with high resolution spectroscopy and enables the discovery and characterization of exoplanets at any position ~1 λ/D away from the star.

A unique addition capability of MODHIS is the ability to perform resolved spectroscopy of previously unresolved worlds. Due to the angular resolution of TMT, planets around nearby (~10 pc) stars residing between 0.01 and 1 au can be spatially resolved for the first time (Figure 6). These planets are bright due to being close to their host stars, but are far enough away that MODHIS can suppress the light of their host stars, unlike current thermal emission studies of close-in planets with current ground-based observatories or with JWST. This allows for a unique sensitivity window for MODHIS to study planets at the boundary of Neptune and Saturn (0.1 $M_{Jup}$).

## 3.3 Exoplanet Masses, Orbits, and Discovery with NIR PRV

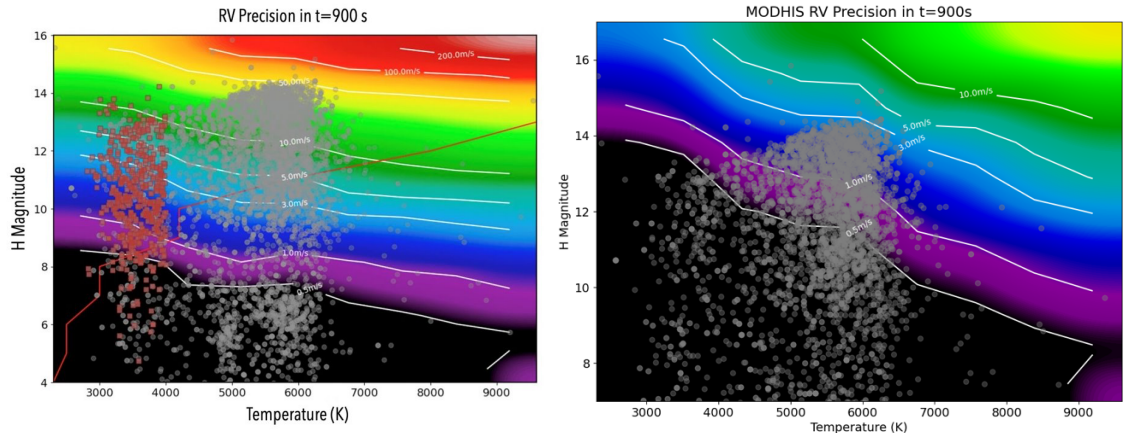

Figure 7. Predicted RV precision with HISPEC (left) and MODHIS (right) as a function of stellar temperature and H band magnitude. Both instruments probe into the m s$^{-1}$ regime for cooler stars, pushing into the often-unexplored region near 3000 K, or the late M dwarf regime. Cool targets such as TRAPPIST-1 are in this spectral bin, showing the great need for PRV capabilities in the NIR where these sources are intrinsically brighter. Contour lines are labeled with the expected RV precision. Grey dots are stars with confirmed planets. Brown dots are TESS objects of interest around stars with T< 4000 K. HISPEC will offer 10 m s$^{-1}$ precision for almost all sources brighter than 14$^{th}$ magnitude. Meanwhile, MODHIS will offer better than 10 m s$^{-1}$ precision on all sources brighter than 16$^{th}$ magnitude. Both capabilities will enable a wide range of scientific inquiry because of the unique high spectral and spatial resolution design of these instruments.

HISPEC and MODHIS offer unique advantages of extreme sensitivity for Precision Radial Velocity measurements (PRV) for cool stars, high spectral and angular resolution and, in some cases, coordinated operation with the visible-light PRV instruments, such as Keck Planet Finder (KPF) in the case of HISPEC.[51] The PRV precision resulting from combining all four wavelength bands will be roughly a factor of two better than that of an individual band, with the exact degree of improvement depending on stellar temperature. An aspect of combining the wavelengths not captured in these figures is the benefit of such a wide wavelength grasp in reducing the effects of stellar variability.[52] Our instrument error budget predicts internal instrumental stability of <0.5 m/s, to be validated by measuring the LFC simultaneously in multiple SMFs. Figure 7 shows HISPEC's (left) and MODHIS's (right) theoretical PRV sensitivity. The practical limit will be set by the effects of telluric absorption, ~1 m/s in a single measurement after post-processing (see below). Figure 8 shows a simulated spectrum from HISPEC (left) and MODHIS (right) of the planet-bearing M8 star Trappist-1 consistent with ~1 m/s accuracy. The sensitivity of HISPEC and MODHIS will enable RV mass measurements for small, temperate, transiting planets orbiting cool stars such as Trappist-1 and other *TESS* candidates. There are currently fewer than 30 confirmed transiting planets orbiting M dwarfs with R < 2R$_\oplus$ and RV masses. None are located in the Habitable Zone (HZ), and there are only two HZ systems orbiting M stars with TTV-determined masses (Trappist-1 and Kepler 138), with many Kepler, K2 or TESS candidates with planet equilibrium temperatures <300 K and R $\leq$ 2R$_\oplus$ awaiting confirmation and characterization. TESS's extended mission will continue to add new objects to this important population. In addition to enabling the follow up of these systems, there are a number of specific NIR PRV science cases uniquely enabled by HISPEC and MODHIS, as described in the next four subsections.

### 3.3.1 Orbital Obliquities of Planets Orbiting Young/Cool Stars

Orbital obliquity, the angle between the stellar spin axis and a planet's angular momentum vector, places crucial constraints on planet formation and migration mechanisms since large mutual inclinations can be highly disruptive in multiple systems.[53,54] The obliquity is measured via the Rossiter- McLaughlin (RM) effect which changes the RV signature during a transit.[55] HISPEC and MODHIS will also be able to determine the obliquity via the "Doppler-shadow" analysis that directly detects the profile variation of the cross-correlation function of the observed spectra.[56] With each observation lasting only a few hours, HISPEC and MODHIS could measure obliquities of planets ranging in size from super-Earths and sub-Neptunes orbiting late M (Teff<4000 K) stars. Only one system cooler than 4000 K, GJ436, currently has a published obliquity.[57] HISPEC and MODHIS will also provide obliquity determinations for planets orbiting young stars. Gaidos et al. [56] recently used Subaru's IRD spectrometer to determine that V1298 Tau b is in a λ ~ 15° prograde orbit, suggesting that the four planets in the system formed in a flat co-planar disk. The statistics of the population of young and mature stars will constrain theories of planet formation and evolution.

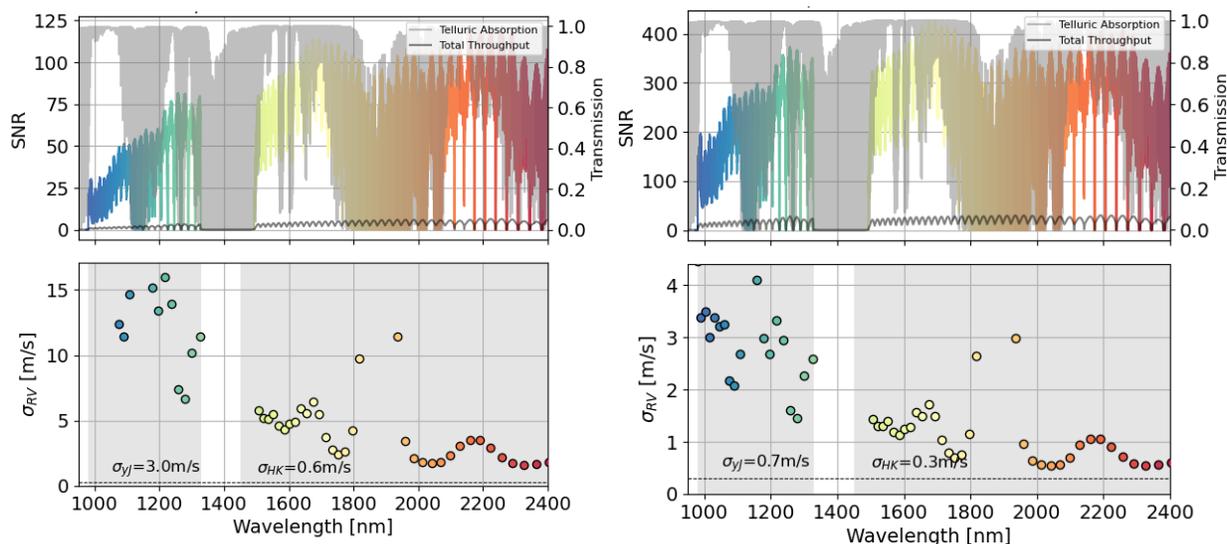

Figure 8. Simulated 30-minute observation of TRAPPIST-1 HISPEC (left) and MOHDIS (right). The top panels show the expected SNR of the source as a function of wavelength while the bottom shows the expected RV precision. While the blue channel of HISPEC only reaches 3 m/s, the H and K band observations provide precision close to below 1 m/s, which is necessary to measure the RV signal for the planets in this system. The expectation for observations of the system with MODHIS is about a factor of 3-4 improvement in precision, which is demonstrated with the simulation shown on the right, where all bands give precision better than 1 m/s.

### 3.3.2 Young Planets

The early evolution of young, still-contracting planets is poorly understood.[58,59] K2 has identified planets transiting young host stars in Young Moving Groups (~50–90 Myr)[60] and open clusters (~600–800 Myr)[61,62] and TESS is discovering many more, e.g., AU Mic b.[52] Among these are hot Jupiters with large predicted Doppler semi-amplitudes of ~10–100 m/s. With the demonstrated 2–4× reduction in the effects of stellar activity relative to visible wavelengths,[63,64] HISPEC and MODHIS will determine the masses of young transiting planets and address the initial density of exoplanets. Recent PRV results for the 20 Myr old star V1298 Tau suggest a surprisingly high initial densities for two of its Jupiter-sized planets[65] compared with theoretical expectations.[66,67,68,69] While additional observations have called this result into question, the fact remains that this is an area of intense observational and theoretical interest which will require extensive observations for its resolution. The accurate determination of the masses and radii of young planets will be critical to understanding the mechanism(s) responsible for the "radius gap".[70,71] Understanding whether or not the population of dense super-Earths found in transit surveys initially formed as rocky bodies or started out as larger, gas rich bodies but lost their $H_2$ rich atmospheres to end up as rocky cores pertains directly to the occurrence rate of Habitable Zone Earth-analogs which are the targets for future direct imaging space missions.[71]

### 3.3.3 Exomoons

Several exomoon candidates have been proposed to date, but none have been confirmed.[72,73] As part of the formation process of exoplanets, exomoons are an important missing piece to better understand planetary systems and can strongly influence the atmospheres of giant exoplanets (e.g., interaction between Io and Jupiter). Icy moons gas giant companions are also interesting objects as they could develop life,[74] and thus constraining the occurrence of such moons is an critical.

In the same way that planets are detected in RV surveys of stars, measuring the radial velocity of planets and brown dwarfs is a promising technique to search for binary planets and exomoons.[75] The mass ratio of moons that form in a circumplanetary disk relative to their planet is expected to be around $10^{-4}$,[76] which is consistent with Ganymede around Jupiter for example. This small mass ratio could explain the lack of current detections.

Using RV measurements from a R=4,000 spectrograph at Keck and the first planetary RV time series,[77] Vanderburg & Rodriguez [78] derived the first exomoon mass upper limits with this technique around the HR 8799 planets. They ruled out Jupiter-mass moons orbiting the 7 Jupiter masses planet HR 8799 c in periods shorter than 1 day. Using KPIC, Ruffio et al. [79] demonstrated a mass ratio sensitivity down to a few percent for satellites orbiting the brown dwarf HR 7672 B. They also explored in detail the prospects The also explored in detail the prospects of exomoon detection with HISPEC and MODHIS around several prime candidate systems, including HR 8799c. They found that planet-sized moons will be detectable with HISPEC, while satellites with similar mass ratio to the large Solar System satellites will be detectable with MODHIS (Figure 5). Batygin & Morbidelli [80] suggests that the moon mass might scale with the planet mass to the 3/2 power. This means that the mass ratios around the larger directly imaged planets and brown dwarfs companion could host significantly larger moons that would be detectable with MODHIS in a few nights of observations. which will be detectable with a few nights of observations with MODHIS.

### 3.3.4 Planet Demographics in Binary Systems

HISPEC and MODHIS's diffraction-limited angular resolution will uniquely enable the study of transiting planets in close binaries (<0.5″) which are typically ignored by seeing-limited instruments.[81] Knowing which star of a pair hosts the planet makes a dramatic difference to the derived planet radius and density[82] and a demographic study will inform how planet formation proceeds in binary systems, a topic of great theoretical interest.[53,83,84]

### 3.4 Spectropolarimetry of Exoplanets with MODHIS[α]

The design of MODHIS is optimized to provide detailed information on the atmospheres and dynamics of a range of astronomical objects. However, there are unique aspects of characterization that cannot be achieved with typical spectroscopic observations alone. Spectral observations of polarized light offer a unique window into a variety of astronomical phenomena, such as the structure of circumstellar or circumplanetary material, the existence and strength of stellar or planetary magnetic fields, the properties of clouds in planetary atmospheres, or potentially the existence of oceans on terrestrial planets. The current design for MODHIS includes a spectropolarimetry mode, in which polarized light can be measured as a function of wavelength in addition to total intensity light. This capability is ***unique*** amongst the first light suite of instruments for all ELTs.

### 3.4.1 Planetary Atmospheres in Polarized Light

Since a polarization signature is commonly produced by light being scattered, detecting polarized light can uniquely probe the properties of the scattering bodies. On an exoplanet, there are a number of potential scattering sources, such as solids that are parts of aerosols or clouds, or liquids on the planetary surface, that could yield a strongly polarized signal. Given the potential of TMT to detect planets smaller than Jupiter in reflected light, having the capability of exploring their atmospheres and surfaces in depth via spectropolarimetry is a unique opportunity afforded by MODHIS. One of the cornerstone science cases for polarimetry for the two decades has been the detection and characterization of atmospheric condensates, namely clouds, hazes, and/or aerosols. The red colors of brown dwarfs and Jovian-mass companions, coupled with often strong signatures of variability, affirm the likely ubiquity of clouds amongst planets like Jupiter.[85] Furthermore, the measurement of a large number of featureless transmission spectra of smaller Neptune-like planets has been attributed to the presence of hazes or aerosols.[5] Indeed, all planets in the Solar System with atmospheres harbor condensates of some variety, spanning a range of compositions and properties. At the same time, with current spectral datasets, clouds remain extremely challenging to characterize and model. Clouds are inherently complex, three-dimensional structures, but tend of be approximated as one dimensional, uniform sources of opacity. Part of the complexity stems from the uncertainty in properties such as the composition of the clouds and the distribution of particle sizes in the clouds. These properties can be explored with polarimetry. Detection of self-luminous objects like brown dwarfs or young Jovian exoplanets with spectropolarimetry will enable the dramatic expansion of current observations with polarized light, which are typically only single wavelength photometric measurement.[86] Spectra in polarized light probe different atmospheric depths, which enables modeling of the structure of the clouds as a function of scale height. Wavelength-dependent data also provides information about grain composition. Additionally, light observed in reflection of sources such as Hot Jupiters could yield upwards of 60% polarization fraction, providing strong constraints on grain properties and cloud compositions.[87,88]

The polarization off of atmospheric molecules (as opposed to condensates) has been explored extensively both for gas giants and for smaller planets. Rayleigh scattering, which we are highly familiar with on Earth, generates among the strongest polarization signals, with models predicting stronger polarization in cloud-free than cloudy

---

[α] A spectropolarimetry mode is not currently planned for HISPEC

atmospheres.[88,89,90,91] For terrestrial planets, both Venus and Earth have been studied extensively in polarized light to understand the potential polarization signal from a similar exoplanet. Observations from satellites or of the Moon have provide a detailed look at "earthshine", or the sunlight scattered off the Earth. Miles-Páez et al. [92] used ground based spectropolarimetry measurements of the earthshine reflection off of the Moon to measure the polarized signal from the Earth across the optical and near-infrared. While the signal is strongest in the optical, the degree of polarization is above 2% across the whole near-infrared, which a very strong signal in the Y-J band from water and $O_2$. Indeed, the polarizing properties of water are extremely strong and well-documented.[93,94] This shows that species such $O_2$ might be more easily found via spectropolarimetry than with traditional absorption line spectroscopy. With MODHIS having the sensitivity to detect these types of planets in transit with NFIRAOS, and potentially via direct imaging with upgraded AO systems like PSI, the capability of spectropolarimetry becomes critical for the goal of characterizing Earth-like worlds. Spectropolarimetry gets us a step further by offering the ability to probe surface features. For example, Gordon et al. [95] explored the wavelength-dependent polarimetric signal from surface features on Earth-like planets, including ocean, sand, and trees. Oceans strongly polarize light, with up to 70% polarization fraction around ~1-1.2 μm. Ice, forests, sand, and grass are also strongly polarizing, often in similar wavelength regimes. However, combinations of surface features can be distinguished by comparing the ratio of various Stokes parameters. While these results are based only on the Earth, they make a strong argument for the use of spectropolarimetry in future missions focused on biosignature work. Thus, MODHIS has a key role to play in the era of characterization of Earth-like planets and the search for life signs on these worlds.

### 3.4.2 Stellar and Planetary Magnetic Fields

Light becomes polarized in the presence of magnetic fields, and thus measurements of polarized light have often been used to infer magnetic field strengths and geometries. Zeeman Doppler imaging takes advantage of this by using spectropolarimetry of rotating stars to map stellar magnetic fields.[96] If star spots with different polarizations are rotating across the disk of the star, there will be variation in the line profile in polarized light. Data collected previously with other spectropolarimeters such as HARPSpol, SPIRou, ESPaDOnS, and others have yielded fascinating insights on the magnetic field topology of a variety of stellar types.[97,98,99,100,101,102] Thus, performing spectropolarimetry is a key science goal for MODHIS. As the only spectropolarimeter available at first like on the ELTs, MODHIS will be unique in its ability to map the magnetic field properties of faint, low mass stars that host terrestrial planets. Given the high occurrence rate of small planets around M-type stars, magnetic activity has been highlighted as a major factor impacting habitability on these worlds.[103,104,105,106] However, we still lack a complete understanding of the origin and structure of these fields, especially on the smallest stars.[107] Zeeman Doppler imaging with MODHIS will offer the first opportunity to map, for example, the magnetic field on systems such as TRAPPIST-1. The second goal of Zeeman Doppler imaging with MODHIS will be to create the first magnetic field maps for gas giant exoplanets. Magnetic fields are found in the gas giants in our Solar System such as Jupiter, but have not been definitively measured on any other exoplanets. Beyond simple detection of magnetic activity, MODHIS will be able to explore the structure of the magnetic field and therefore provide hints about the interiors of these planets. This will also shed light on possible interactions between the magnetic field of the planet and orbiting moons, such as those between Jupiter and Io that give rise to aurora on Jupiter.[108]

### 3.5 Physics at the Bottom of the Main Sequence

HISPEC and MODHIS have the potential to revolutionize our understanding of the detailed properties of the coldest stars and brown dwarfs, the late-M, L, T, and Y "ultracool" dwarfs.[109] These objects, like their warm exoplanet analogs, have cool atmospheres whose spectral energy distributions, shaped by strong atomic and molecular absorption, peak in the infrared. And, as intrinsically faint sources ($L \leq 10^{-4}$ $L_\odot$), the accessible population is better suited to individual targeting than broad sky surveys and are thus ideal targets for HISPEC and MODHIS spectroscopy.

The combination of broad wavelength coverage and high-resolution spectroscopy provided by HISPEC and MODHIS will enable novel investigations into the systemic and atmospheric properties of these objects. The availability of thousands of resolved molecular features across the 1-2.5 μm region will enable PRV measurements ($\sigma_{RV} < 100$ m/s) that match the astrometric precision of *Gaia*, providing 3D velocity and 6D coordinate vector measurements for population dynamics studies,[110] cluster membership,[111] and the identification of so-called "fly-by" stars that have/will have close passage of the Sun.[112] PRV measurements will also enable the first detection of gravitational redshift from ultracool dwarfs (of order 300-700 m/s), whose distinct mass/radius dependence compared to surface gravity ($RV_z \propto M/R$, $g \propto M/R^2$) provides an opportunity for direct measurement of the substellar, semi-degenerate mass/radius relationship.[113] PRV monitoring of resolved (exploiting the high spatial resolution of the Laser Guide Star AO feed) and unresolved ultracool dwarf systems (periods of weeks to years) will provide orbit maps that sample the underlying separation, mass ratio, and eccentricity distributions of these systems, critical for testing low-mass star and brown dwarf formation theories.[114] HISPEC and MODHIS provide a particularly unique opportunity for late-M/L plus T dwarf "spectral binaries" for which both primary and secondary RVs can be captured simultaneously from different parts of the spectrum.[115,116]

MODHIS spectroscopy will also enable rotational velocity studies of low-mass stars and brown dwarfs in the field and in clusters of various ages, providing new insights into angular momentum evolution as a function of mass and age in the regime of fully convective interiors and neutral atmospheres.[117] Such measurements are particularly valuable for samples of photometric variability ultracool dwarfs, increasingly available through facilities such as TESS,[118] SPECULOOS,[119] and PINES,[120] as they enable statistical radius measurements and studies of latitudinal structure.[121,122] In addition, by comparing Zeeman line broadening or splitting of magnetic-sensitive features such as K I (1.24 μm) and FeH (0.98 μm, 1.2 μm, and 1.5 μm) to non-sensitive features, we will be able to directly examine the correlation between magnetic field strength and rotation,[123,124] and potentially map magnetic structures (Zeeman Doppler Imaging; §3.4.2),[125,126] as empirical tests for dynamo models in the fully-convective regime.[127,128] Spectral monitoring of ultracool stars and brown dwarfs will also provide insights into the 3D structures and dynamics of condensate clouds, which form in both ultracool dwarf and giant exoplanet atmospheres.[129] The continuum absorption and scattering opacity of condensates mutes atomic and molecular features, and when these features are monitored in velocity space (line profile variability) it is possible to generate a Doppler imaging map of cloud structures.[16] Concurrent monitoring of regions of high and low atmospheric opacity, which sample the upper and lower atmospheres respectively, can be used to build a combined vertical and azimuthal map of cloud structures that can be compared to 3D atmosphere models.[130,131,132]

## 3.6 The Galactic Center

HISPEC and MODHIS will be transformative for science at the Galactic Center (GC). While astrometric precision has improved with the development of better AO systems and instruments like GRAVITY,[133] RV precision has lagged. With the next generation of telescopes like TMT, we expect the improvements in astrometry to be only a factor of ~3 compared to Keck. In contrast, high spectral resolution and sensitivity from HISPEC and MODHIS can improve RV precision by a factor of 100 (from 10 km/s to 100 m/s) compared to current AO integral-field spectrographs used at the Galactic center. This two orders-of-magnitude improvement will open a new era of investigation of the physics and astrophysics of supermassive black holes. Figure 9 demonstrates the level of improvement expected in the next few years with HISPEC. MODHIS will offer similar RV precision, but for much fainter targets in the GC region.

Although the Standard Model of particle physics does not predict its dependence on space and time, higher dimensional theories often predict variation of the fundamental physical constants. If observed, this will be an unambiguous signal of new physics.[134] A variation of the fine structure constant α was claimed to be observed,[135,136] but is still controversial.[137] Spectroscopy of stars near Sgr A* enables us to measure α in extreme and completely unexplored gravity environments.[138] HISPEC and MODHIS will provide >100× stronger constraint on the variation of α by observing metal absorption lines in late-type stellar spectra ($10^4$–$10^5$ more sensitive to the variation of α than H and He lines). By surveying a number of different stars in this region, it will be possible to statistically probe potential variation on α. This will be started with HISPEC for the brighter stars and followed by MODHIS for fainter objects.

HISPEC and MODHIS will lead the next era of studies about the nature of gravity at the GC. Black holes (BHs) represent the intersection between gravity and quantum mechanics, so they are ideal objects to test the limits of both theories and potentially reveal new physics. Measurement of the spin of Sgr A* would complete the supermassive BH description. The "no hair" theorem states that mass, spin, and charge are the only necessary parameters and the only ones observable from outside a BH. The spin is the only remaining parameter to measure for Sgr A*. While there have been spin estimates,[139,140] they all require significant assumptions. In contrast, the equations of motion are well known for stars in orbit around Sgr A*, and their very precise measurements can yield the drag on space-time caused by the spinning BH.[141,142] This requires spatially resolved stellar RVs with uncertainties of <100 m/s,[142] which HISPEC and MODHIS will easily achieve.

There is a large debate around the existence of an intermediate-mass BH (IMBH) at the GC, which could be constrained by the measurements of the S-stars' orbits and from the measurement of the acceleration of stars at larger distances. GRAVITY has performed precise astrometry of S0-2 which excludes the existence of any IMBH with ≳1000 $M_\odot$ in the central 0.1 arcsecond. The allowed mass range of an IMBH is 100–1000 $M_\odot$ inside or just outside of the orbit of S0-2,[133] HISPEC and MODHIS will reduce S-stars' RV uncertainties from a few km/s to <100 m/s, providing a stronger limit on an IMBH orbiting Sgr A*. The instruments will also enable stellar dynamical searches for IMBH companions at further distances such as within the IRS 13 group of stars at 0.1 pc away from the black hole. There are suggestions based on ALMA observations of gas that there may be a 10,000 $M_\odot$ IMBH at the center of this system.[143] To definitively show whether this is a black hole requires detecting acceleration of the nearby stars from this IMBH candidate. Only instruments like HISPEC and MODHIS can achieve the acceleration sensitivity requirement of 0.01 km/s/yr for stars within 0.2 arcsecond of the IMBH candidate.

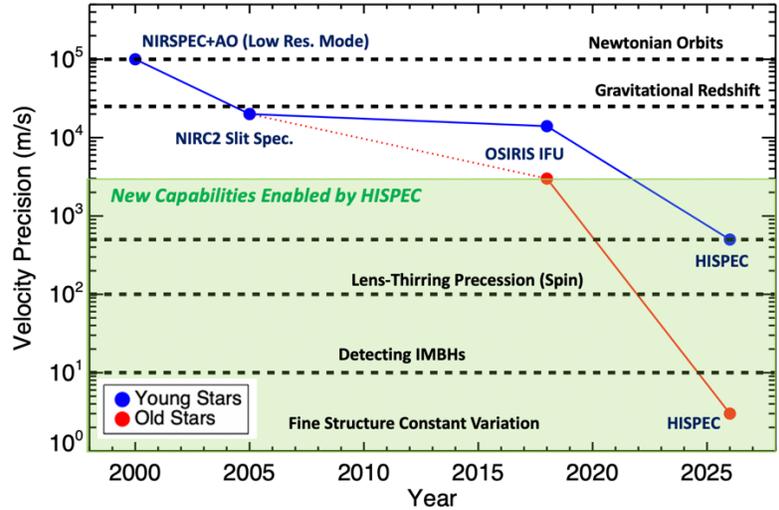

Figure 9. Change in time of RV precision achievable on sources in the Galactic Center using Keck instrumentation. The current best precision is offered by the OSIRIS integral field spectrograph, which has R~4000. By increasing the spectral resolution and achieving high sensitivity, HISPEC will vastly improve the precision on RV measurements of stars in this region, offering new scientific opportunities. MODHIS will offer similar precision, but target much fainter sources, which will enable a broader survey of the region at very high spectral resolution.

## 4. CONCLUSIONS

The combination of high spatial and high spectral resolution has the potential to revolutionize several areas of observational research in the next several years. With HISPEC coming online in 2026 and MODHIS in the early 2030s, both instruments have the potential for synergistic operation not only with JWST, but also with Roman and the future Habitable Worlds Observatory. Additional science cases for both instruments not included in this proceeding include studies of the kinematics and compositions of young stars in star forming regions, probing peculiar abundances of stars in the Galactic bulge and halo, measuring the precise velocities of faint dwarf galaxy members, and studying the sphere of influence around AGN in nearby galaxies. These science cases will be documented in the near future in the detailed science case documents for both instruments.

## REFERENCES

[1] Dodson-Robinson, S.E., Veras, D., Ford, E.B., and Beichman, C.A., "The Formation Mechanism of Gas Giants on Wide Orbits", ApJ, 707, 79-88 (2009)
[2] Beatty, T.G., Marley, M.S., Gaudi, S.B., Colón, K.D., Fortney, J.J., and Showman, A.P., "Spitzer Phase Curves of KELT-1b and the Signatures of Nightside Clouds in Thermal Phase Observations", AJ, 158, 166 (2019)
[3] Gao, P., Thorngren, D.P., Lee, G.K.H., Fortney, J.J., Morley, C.V., Wakeford, H.R., Powell, D.K., Stevenson, K.B., and Zhang, X., "Aerosol composition of hot giant exoplanets dominated by silicates and hydrocarbon hazes", Nature Astronomy, 4, 951–956 (2020)
[4] Madhusudhan, N., "Exoplanetary Atmospheres: Key Insights, Challenges, and Prospects", ARA&A, 57, 617-663 (2019)

[5] Sing, D.K., Fortney, J.J., Nikolov, N., Wakeford, H.R., Kataria, T., Evans, T.M., Aigrain, S., Ballester, G.E., Burrows, A.S., Deming, D., Désert, J.-M., Gibson, N.P., Henry, G.W., Huitson, C.M., Knutson, H.A., Lecavelier Des Etangs, A., Pont, F., Showman, A.P., Vidal-Madjar, A., Williamson, M.H., and Wilson, P.A. "A continuum from clear to cloudy hot-Jupiter exoplanets without primordial water depletion", Nature, 529, 59–62, (2016)

[6] Rajan, A., Rameau, J., DeRosa, R.J., Marley, M.S., Graham, J.R., Macintosh, B., Marois, C., Morley, C., Patience, J., Pueyo, L., Saumon, D., Ward-Duong, K., Ammons, S.M., Arriaga, P., Bailey, V.P., Barman, T., Bulger, J., Burrows, A.S., Chilcote, J., Cotten, T., Czekala, I., Doyon, R., Duchêne. G., Esposito, T.M., Fitzgerald, M.P., Follette, K.B., Fortney, J.J., Goodsell, S.J., Greenbaum, A.Z., Hibon, P., Hung, L.-W., Ingraham, P., Johnson-Groh, M., Kalas, P., Konopacky, Q., Lafrenière, D., Larkin, J.E., Maire, J., Marchis, F., Metchev, S., Millar-Blanchaer, M.A., Morzinski, K.M., Nielsen, E.L., Oppenheimer, R., Palmer, D., I. Patel, R.I., Perrin, M., Poyneer, L., Rantakyrö, F.T., Ruffio, J.-B., Savransky, D., Schneider, A.C., Sivaramakrishnan, A., Song, I., Soummer, R., Thomas, S., Vasisht, G., Wallace, J.K., Wang, J.J., Wiktorowicz, S., and Wolff, S., "Characterizing 51 Eri b from 1 to 5 µm: A Partly Cloudy Exoplanet" AJ, 154,10, (2017)

[7] Miles, B.E., Biller, B.A., Patapis, P., Worthen, K., Rickman, E., Hoch, K.K.W., Skemer, A., Perrin, M.D.., Whiteford, N., Chen, C.H., Sargent, B., Mukherjee, S., Morley, C.V., Moran, S.E., Bonnefoy, M., Petrus, S., Carter, A.L., Choquet, E., Hinkley, S., Ward-Duong, K., Leisenring, J.M., Millar-Blanchaer, M.A., Pueyo, L., Ray, S., Sallum, S., Stapelfeldt, K.R., Stone, J.M., Wang, J.J., Absil, O., Balmer, W.O., Boccaletti, A., Bonavita, M., Booth, M., Bowler, B.P., Chauvin, G., Christiaens, V., Currie, T., Danielski, C., Fortney, J.J., Girard, J.H., Grady, C.A., Greenbaum, A.Z., Henning, Th., Hines, D.C., Janson, M., Kalas, P., Kammerer, J., Kennedy, G.M., Kenworthy, M.A., Kervella, P., Lagage, P.-O., Lew, B.W.P., Liu, M.C., Macintosh, B., Marino, S., Marley, M.S., Marois, C., Matthews, E.C., Matthews, B.C., Mawet, D., McElwain, M.W., Metchev, S., Meyer, M.R., Molliere, P., Pantin, E., Quirrenbach, A., Rebollido, I., Ren, B.B., Schneider, G., Vasist, M., Wyatt, M.C., Zhou, Y., Briesemeister, Z.W., Bryan, M.L., Calissendorff, P., Cantalloube, F., Cugno, G., De Furio, M., Dupuy, T.J., Factor, S.M., Faherty, J.K., Fitzgerald, M.P., Franson, K.,Gonzales, E.c., Hood, C.E., Howe, A.R., Kraus, A.L., Kuzuhara, M., Lagrange, A.-M., Lawson, K., Lazzoni, C., Liu, P., Llop-Sayson, J., Lloyd, J.P., Martinez, R.A., Mazoyer, J., Quanz, S.P., Redai, J.A., Samland, M., Schlieder, J.E., Tamura, M., Tan, X., Uyama, T., Vigan, A., Vos, J.M., Wagner, K., Wolff, S.G., Ygouf, M., Zhang, X., Zhang, K., Zhang, Z., "The JWST Early-release Science Program for Direct Observations of Exoplanetary Systems II: A 1 to 20 µm Spectrum of the Planetary-mass Companion VHS 1256-1257 b", ApJ, 946, 6 (2023)

[8] Alderson, L., Wakeford, H.R., Alam, M.K., Batalha, N.E., Lothringer, J.D., Adams Redai, J.,; Barat, S., Brande, J., Damiano, M., Daylan, T., Espinoza, N., Flagg, L., Goyal, J.M., Grant, D., Hu, R., Inglis, J., Lee, E.K.H., Mikal-Evans, T., Ramos-Rosado, L., Roy, P.-A., Wallack, N.L., Batalha, N.M., Bean, J.L., Benneke, B., Berta-Thompson, Z.K., Carter, A.L., Changeat, Q., Colón, K.D., Crossfield, I.J.M., Désert, J.-M., Foreman-Mackey, D., Gibson, N.P., Kreidberg, L., Line, M.R., López-Morales, M., Molaverdikhani, K., Moran, S.E., Morello, G., Moses, J.I., Mukherjee, S., Schlawin, E., Sing, D.K., Stevenson, K.B., Taylor, J., Aggarwal, K., Ahrer, E.-M., Allen, N.H., Barstow, J.K., Bell, T.J., Blecic, J., Casewell, S.L., Chubb, K.L., Crouzet, N., Cubillos, P.E., Decin, L., Feinstein, A.D., Fortney, J.J., Harrington, J., Heng, K., Iro, N., Kempton, E.M-R., Kirk, J., Knutson, H.A., Krick, J., Leconte, J., Lendl, M., MacDonald, R.J., Mancini, L., Mansfield, M., May, E.M., Mayne, N.J., Miguel, Y., Nikolov, N.K., Ohno, K., Palle, E., Parmentier, V., Petit dit de la Roche, D.J.M., Piaulet, C., Powell, D., Rackham, B.V., Redfield, S., Rogers, L.K., Rustamkulov, Z., Tan, X., Tremblin, P., Tsai, S.-M., Turner, J.D., de Val-Borro, M., Venot, O., Welbanks, L., Wheatley, P.J., Zhang, X, "Early Release Science of the exoplanet WASP-39b with JWST NIRSpec G395H", Nature, 614, 664-669 (2023)

[9] Kreidberg, L., Bean, J.L., Désert, J.-M., Benneke, B., Deming, D., Stevenson, K.B., Seager, S, Berta-Thompson,, Z. Seifahrt, A, and Homeier, D., "Clouds in the atmosphere of the super-Earth exoplanet GJ1214b" Nature, 505, 69–72, (2014)

[10] Hood, C.E., Fortney, J.J., Line, M.R., Martin, E.C., Morley, C.V., Birkby, J.L., Rustamkulov, Z., Lupu, R.E., and Freedman, R.S., "Prospects for Characterizing the Haziest Sub-Neptune Exoplanets with High-resolution Spectroscopy". AJ, 160, 198, (2020)

[11] Gandhi, S., Brogi, M., and Webb, R.K, "Seeing above the clouds with high-resolution spectroscopy", MNRAS, 498, 194–204 (2020)

[12] Snellen, I. A. G., de Kok, R. J., de Mooij, E. J. W., and Albrecht, S., "The orbital motion, absolute mass and high-altitude winds of exoplanet HD209458b", Nature, 465, 1049–1051 (2010)

[13] Brogi, M., Snellen, I. A. G., de Kok, R. J., Albrecht, S., Birkby, J., and de Mooij, E. J. W., "The signature of orbital motion from the dayside of the planet τ Boötis b", Nature, 486, 502–504 (2012)

[14] Birkby, J. L., de Kok, R. J., Brogi, M., de Mooij, E. J. W., Schwarz, H., Albrecht, S., and Snellen, I. A. G., "Detection of water absorption in the day side atmosphere of HD 189733 b using ground-based high- resolution spectroscopy at 3.2 μm", MNRAS, 436, L35–L39 (2013)

[15] Snellen, I. A. G., Brandl, B. R., de Kok, R. J., Brogi, M., Birkby, J., and Schwarz, H. "Fast spin of the young extrasolar planet β Pictoris b". Nature, 509, 63–65 (2014)

[16] Crossfield, I. J. M., "Doppler imaging of exoplanets and brown dwarfs", A&A, 566, A130 (2014)

[17] Snellen, I., de Kok, R., Birkby, J.L., Brandl, B., Brogi, M., Keller, C., Kenworthy, M., Schwarz, H., and Stuik, R., "Combining high-dispersion spectroscopy with high contrast imaging: Probing rocky planets around our nearest neighbors", A&A, 576, A59 (2015)

[18] Brogi, M., de Kok, R. J., Albrecht, S., Snellen, I. A. G., Birkby, J. L., and Schwarz, H., "Rotation and Winds of Exoplanet HD 189733 b Measured with High-dispersion Transmission Spectroscopy" ApJ, 817, 106, (2016)

[19] Birkby, J.L., "Exoplanet Atmospheres at High Spectral Resolution", arXiv:1806.04617 (2018)

[20] Pelletier, S., Benneke, B., Darveau-Bernier, A., Boucher, A., Cook, N.J., Pilaulet, C., Coulombe, L.-P., Artigau, E., Lafrenière, D., Delisle, S., Allart, R., Doyon, R., Donati, J.-F., Fouqué, P., Moutou, C., Cadieux, C., Delfosse, X., Hébrard, G., Martins, J.H.C., Martioli, E., and Vandal, T., "Where Is the Water? Jupiter-like C/H Ratio but Strong H2O Depletion Found on τ Boötis b Using SPIRou" AJ, 162, 73 (2021)

[21] Line, M.R., Brogi, M., Bean, J.L., Gandhi, S., Zalesky, J., Parmentier, V., Smith, P., Mace, G.N., Mansfield, M., Kempton, E.M.-R., Fortney, J.J., Shkolnik, E., Patience, J., Rauscher, E., Désert, J.-M., and Wardenier, J.P., "A solar C/O and sub-solar metallicity in a hot Jupiter atmosphere", Nature, 598, 580–584 (2021)

[22] Brogi, M., Line, M., Bean, J,. Désert, J. M., and Schwarz, H., "A Framework to Combine Low- and High-resolution Spectroscopy for the Atmospheres of Transiting Exoplanets", ApJ, 839, L2 (2017)

[23] Brogi, M. and Line, M.R., "Retrieving Temperatures and Abundances of Exoplanet Atmospheres with High-resolution Cross-correlation Spectroscopy", AJ, 157, 114 (2019)

[24] Mawet, D., Fitzgerald, M.P., Konopacky, Q., Jovanovic, N., Baker, A., Beichman, C., Bertz, R., Dekany, R., Fucik, J., Roberts, M., Porter, M., Pahuja, R., Ruane, G., Leifer, S., Halverson, S., Gibbs, A., Johnson, C., Kress, E., Magnone, K., Sohn, J.M., Wang, E., Brown, A., Maire, J., Sappey, B., Andersen, D., Terada, H., Kassis, M., Artigau, E., Benneke, B., Doyon, R., Kotani, T., Tamura, M., Beatty, T., Plavchan, P., Do, T., Nishiyama, S., Wang, J., Wang, J., "Fiber-fed high-resolution infrared spectroscopy at the diffraction limit with Keck-HISPEC and TMT-MODHIS: status update", Proc. SPIE (2017)

[25] Echeverri, D., Ruane, G., Calvin, B., Jovanovic, N., Delorme, J.-R., Wang, J., Millar-Blanchaer, M., Mawet, D., Serabyn, E., Wallace, J.K., and Martin, S., "Detecting and characterizing close-in exoplanets with vortex fiber nulling", Proc. SPIE, 11446 (2020)

[26] Echeverri, D., Ruane, G., Jovanovic, N., Delorme, J.-R., Wang, J., Millar-Blanchaer, M.A., Xuan, J., Toman, K., and Mawet, D., "Broadband vortex fiber nulling: high-dispersion exoplanet science at the diffraction limit", Proc. SPIE, 11823 (2021)

[27] Delorme, J.-R., Jovanovic, N., Echeverri, D., Mawet, D., Wallace, J.K., Bartos, R.D., Cetre, S., Wizinowich, P., Ragland, S., Lilley, S., Wetherell, E., Doppmann, G., Wang, J.J., Morris, E.C., Ruffio, J.-B., Martin, E.C., Fitzgerald, M.P., Ruane, G., Schofield, T., Suominen, N., Calvin, B., Wang, E., Magnone, K., Johnson, C., Sohn, J.M., López, R.A., Bond, C.Z., Pezzato, J., Sayson, J.L., Chun, M., Skemer, A.J., "Keck Planet Imager and Characterizer: a dedicated single-mode fiber injection unit for high-resolution exoplanet spectroscopy", JATIS, 7, 5006

[28] Yi, X., Vahala, K., Li, J., Diddams, S., Ycas, S., Plavchan, P., Leifer, S., Sandhu, J., Vasisht, G., Chen, P., Gao, P., Gagne, J., Furlan, E., Bottom, M., Martin, E. C., Fitzgerald, M. P., Doppmann, G., and Beichman, C., "Demonstration of a near-IR line-referenced electro-optical laser frequency comb for precision radial velocity measurements in astronomy", Nature Communications, 7, 10436 (2016)

[29] Crane, J., Herriot, G., Andersen, D., Atwood, J., Byrnes, P., Densmore, A., Dunn, J., Fitzsimmons, J., Hardy, T., Hoff, B., Jackson, K., Kerley, D., Lardière, O., Smith, M., Stocks, J., Véran, J.-P., Boyer, C., Wang, L., Trancho, G., and Trubey, M., "NFIRAOS adaptive optics for the Thirty Meter Telescope", Proc SPIE, 10703 (2018)

[30] Atwood, J., Hoff, B., Andersen, D., Crane, J., and Dunn, J., "InfraRed Imaging Spectrograph (IRIS) on TMT: OIWFS opto-mechanical design update", Proc. SPIE, 11448 (2020)

[31] Surya, A., Zonca, A., Rundquist, N.-E., Wright, S.A., Walth, G.L., Anderson, D.R., Chapin, E.L., Chisholm, E.M., Do, T., Dunn, J.S., Gillies, K., Hayano, Y., Johnson, C.A., Kupke, R., Larkin, J.E., Nakamoto, T., Riddle, R., Smith, R., Suzuki, R., Sohn, J.M., Weber, R., Weiss, J.L., Zhang, K., "The Infrared Imaging Spectrograph (IRIS) for TMT: final design development of the data reduction system", Proc. SPIE, 11452 (2020)

[32] Gibbs, A., Fitzgerald, M., Mawet, D., Konopacky, Q., Baker, A., Benneke, B., Echeverri, D., Hillman, S., Millar-Blanchaer, M.A., Pelletier, S., Perera, S., Sappey, B., Wang, J., "Echelle simulation for the High-resolution Infrared Spectrograph for Exoplanet Characterization (HISPEC) at Keck", Proc. SPIE, 12184 (2022)

[33] Stevenson, K.B., Désert, J.-M., Line, M.R., Bean, J.L., Fortney, J.J., Showman, A.P., Kataria, T., Kreidberg, L., McCullough, P.R., Henry, G.W., Charbonneau, D., Burrows, A., Seager, S., Madhusudhan, N., Williamson, M.H., Homeier, D., "Thermal structure of an exoplanet atmosphere from phase-resolved emission spectroscopy", Science, 346, 838-841 (2014)

[34] Mordasini, C., van Boekel, R., Mollière, P., Henning, Th., and Benneke, B., "The Imprint of Exoplanet Formation History on Observable Present-day Spectra of Hot Jupiters", ApJ, 832, 41 (2016)

[35] Madhusudhan, N., Bitsch, B., Johansen, A., and Eriksson, L., "Atmospheric signatures of giant exoplanet formation by pebble accretion", MNRAS, 468, 4102-4115 (2017)

[36] Madhusudhan, N., Piette, A.A.A., and Constantinou, S., "Habitability and Biosignatures of Hycean Worlds", ApJ, 918, 1 (2021)

[37] Rackham, B.V., Apai, D., and Giampapa, M.S., "The Transit Light Source Effect: False Spectral Features and Incorrect Densities for M-dwarf Transiting Planets", ApJ, 853, 122, (2018)

[38] Fulton B.J., Petigura, E.A., Howard, A.W., Isaacson, H., Marcy, G.W., Cargile, P.A., Hebb, L., Weiss, L.M., Johnson, J.A., Morton, T.D., Sinukoff, E., Crossfield, I.J.M., and Hirsch, L.A., "The California-Kepler Survey. III. A Gap in the Radius Distribution of Small Planets", AJ, 154, 109 (2017)

[39] Hoeijmakers, H.J., Schwarz, H., Snellen, I.A.G., de Kok, R.J., Bonnefoy, M., Chauvin, G., Lagrange, A.-M., and Girard, J.H., "Medium-resolution integral-field spectroscopy for high-contrast exoplanet imaging. Molecule maps of the β Pictoris system with SINFONI", A&A, 617, A144 (2018)

[40] Bowler, B., Sallum, S., Boss, A., Brandt, T., Briesemeister, Z., Bryan, M., Crepp, J., Currie, T., Fortney, J., Girard, J., Jensen-Clem, R., Kama, M., Kraus, A., Konopacky, Q., Liu, M., Marley, M., Mawet, D., Meshkat, T., Meyer, M., Morley, C., Skemer, A., Wang, J., Wu, Y.-L., Close, L., Marois, C., and Nielsen, E., "The Demographics and Atmospheres of Giant Planets with the ELTs" BAAS Astro2020 White Papers, 51, 496, (May 2019)

[41] Wang, J., Mawet, D., Ruane, G., Hu, R., and Benneke, B., "Observing Exoplanets with High Dispersion Coronagraphy. I. The Scientific Potential of Current and Next-generation Large Ground and Space Telescopes", AJ, 153, 183 (2017)

[42] Mawet, D., Ruane, G., Xuan, W., Echeverri, D., Klimovich, N., Randolph, M., Fucik, J., Wallace, J.K., Wang, J., Vasisht, G., Dekany, R., Mennesson, B., Choquet, E., Delorme, J.-R., and Serabyn, E., "Observing Exoplanets with High-dispersion Coronagraphy. II. Demonstration of an Active Single-mode Fiber Injection Unit", ApJ, 838, 92(2017)

[43] National Academies of Sciences, Engineering, and Medicine. "Pathways to Discovery in Astronomy and Astrophysics for the 2020s", The National Academies Press, Washington, DC (2021)

[44] Fernandes, R.B., Mulders, G.D., Pascucci, I., Mordasini, C., and Emsenhuber, A., "Hints for a Turnover at the Snow Line in the Giant Planet Occurrence Rate", ApJ, 874, 81 (2019)

[45] Rosenthal, L.J., Fulton, B.J., Hirsch, L.A., Isaacson, H.T., Howard, A.W., Dedrick, C.M., Sherstyuk, I.A., Blunt, S.C., Petigura, E.A., Knutson, H.A., Behmard, A., Chontos, A., Crepp, J.R, Crossfield, I.J.M, Dalba, P.A., Fischer, D.A., Henry, G.W., Kane, S.R., Kosiarek, M., Marcy, G.W., Rubenzahl, R.A., Weiss, L.M., and Wright, J.T., "The California Legacy Survey. I. A Catalog of 178 Planets from Precision Radial Velocity Monitoring of 719 Nearby Stars over Three Decades", ApJS, 255, 8 (2021)

[46] Cridland, A.J., van Dishoeck, E.F., Alessi, M., and Pudritz, R.E., "Connecting planet formation and astrochemistry. C/Os and N/Os of warm giant planets and Jupiter analogues", A&A, 642, A229 (2020)

[47] Batygin, K., "On the Terminal Rotation Rates of Giant Planets", AJ, 155, 178 (2018)

[48] Perryman, M., Hartman, J., Bakos, G.A., and Lindegren, L., "Astrometric Exoplanet Detection with Gaia", ApJ, 797, 14 (2014)

[49] Ruane, G., Wang, J., Mawet, D., Jovanovic, N., Delorme, J.-R., Mennesson, B., and Wallace, J. K., "Efficient Spectroscopy of Exoplanets at Small Angular Separations with Vortex Fiber Nulling", ApJ, 867, 143 (2018)

[50] Echeverri, D., Ruane, G., Jovanovic, N., Mawet, D., and Levraud, N., "Vortex fiber nulling for exoplanet observations I Experimental demonstration in monochromatic light", Optics Letters, 44, 2204 (2019)

[51] Gibson, S.R., Howard, A.W., Roy, A., Smith, C., Halverson, S., Edelstein, J., Kassis, M., Wishnow, E.H., Raffanti, M., Allen, S., Chin, J., Coutts, D., Cowley, D., Curtis, J., Deich, W., Feger, T., Finstad, D., Gurevich, Y., Ishikawa, Y., James, E., Jhoti, E., Lanclos, K., Lilley, S., Miller, T., Milner, S., Payne, T., Rider, K., Rockosi, C., Sandford,

D., Schwab, C., Seifahrt, A., Sirk, M.M., Smith, R., Stuermer, J., Weisfeiler, W., Wilcox, M., Vandenberg, A., and Wizinowich, P., "Keck Planet Finder: preliminary design", Proc. SPIE, 107025 (2018)

[52] Plavchan, P., Barclay, T., Gagné, J., Gao, P., Cale, B., Matzko, W., Dragomir, D. Quinn, S., Feliz, D., Stassun, K., Crossfield, I.J.M., Berardo, D.A., Latham, D.W., Tieu, B., Anglada-Escudé, G., Ricker, G., Vanderspek, R., Seager, S., Winn, J.N., Jenkins, J.M., Rinehart, S., Krishnamurthy, A., Dynes, S., Doty, J., Adams, F., Afanasev, D.A., Beichman, Chas, Bottom, Mike, Bowler, Brendan P., Brinkworth, Carolyn, Brown, Carolyn J., Cancino, Andrew, Ciardi, D.R., Clampin, M., Clark, .T., Collins, K., Davison, C., Foreman-Mackey, D., Furlan, E., Gaidos, E.J., Geneser, C., Giddens, F., Gilbert, E., Hall, R., Hellier, C., Henry, T., Horner, J., Howard, A.W., Huang, C., Huber, J., Kane, S.R., Kenworthy, M., Kielkopf, J., Kipping, D., Klenke, C., Kruse, E., Latouf, N., Lowrance, P., Mennesson, B., Mengel, M., Mills, S.M., Morton, T., Narita, N., Newton, E., Nishimoto, A., Okumura, J., Palle, E., Pepper, J., Quintana, E.V., Roberge, A., Roccatagliata, V., Schlieder, J.E., Tanner, A., Teske, J., Tinney, C.G., Vanderburg, A., von Braun, K., Walp, B., Wang, J., Wang, S.X., Weigand, D. White, R., Wittenmyer, R.A., Wright, D.J., Youngblood, A., Zhang, H., Zilberman, P., "A planet within the debris disk around the pre-main-sequence star AU Microscopii", Nature, 582, 497-500 (2020)

[53] Winn, J.N. and Fabrycky, D.C., "The Occurrence and Architecture of Exoplanetary Systems", ARA&A, 53, 409–447 (2015)

[54] Ormel, C.W., Liu, B., and Schoonenberg, D., "Formation of TRAPPIST-1 and other compact systems", A&A, 604, A1 (2017)

[55] Triaud, A.H.M.J., "The Rossiter-McLaughlin Effect in Exoplanet Research", Handbook of Exoplanets, ed. H. Deeg & J.A. Belmonte, Springer Nature, 2 (2018)

[56] Gaidos, E., Hirano, T., Beichman, C., Livingston, J., Harakawa, H., Hodapp, K.W., Ishizuka, M., Jacobson, S., Konishi, M., Kotani, T., Kudo, T., Kurokawa, T., Kuzuhara, M., Nishikawa, J., Omiya, M., Serizawa, T., Tamura, M., Ueda, A., and Vievard, S., "Zodiacal exoplanets in time - XIII. Planet orbits and atmospheres in the V1298 Tau system, a keystone in studies of early planetary evolution", MNRAS, 509, 2969–2978 (2022)

[57] Bourrier, V., Lovis, C., Beust, H., Ehrenreich, D., Henry, G.W., Astudillo-Defru,N., Allart, R., Bonfils, X., Ségransan, D., Delfosse, X., Cegla, H.M., Wyttenbach, A., Heng, K., Lavie, B., and Pepe, F. "Orbital misalignment of the Neptune-mass exoplanet GJ 436b with the spin of its cool star", Nature, 553, 477–480 (2018)

[58] Fortney, J.J., Marley, M.S., Saumon, D., and Lodders. K., "Synthetic Spectra and Colors of Young Giant Planet Atmospheres: Effects of Initial Conditions and Atmospheric Metallicity", ApJ, 683, 1104– 1116 (2008)

[59] Spiegel, D.S. and Burrows, A., "Thermal Processes Governing Hot-Jupiter Radii", ApJ 772, 76 (2013)

[60] David, T.J., Mamajek, E.E., Vanderburg, A., Schlieder, J.E., Bristow, M., Petigura, E.A., Ciardi, D.R., Crossfield, I.J.M., Isaacson, H.T., Cody, A.M., Stauffer, J.R., Hillenbrand, L.A., Bieryla, A., Latham, D.W., Fulton, B.J., Rebull, L.M., Beichman, C., Gonzales, E.J., Hirsch, L.A., Howard, A.W., Vasisht, G., and Ygouf, M., "Discovery of a Transiting Adolescent Sub-Neptune Exoplanet with K2" AJ, 156, 302 (2018)

[61] Mann, A.W., Gaidos, E., Vanderburg, A., Rizzuto, A.C., Ansdell, M., Medina, J.V., Mace, G.N., Kraus, A.L., and Sokal, K.R., "Zodiacal Exoplanets in Time (ZEIT). IV. Seven Transiting Planets in the Praesepe Cluster", AJ, 153, 64 (2017)

[62] Ciardi, D.R., Crossfield, I.J.M., Feinstein, A.D., Schlieder, J.E., Petigura, E.A., David, T.J., Bristow, M., Patel, R.I., Arnold, L., Benneke, B., Christiansen, J.L., Dressing, Co.D., Fulton, B.J., Howard, A.W., Isaacson, H., Sinukoff, E., and Thackeray, B., "K2-136: A Binary System in the Hyades Cluster Hosting a Neptune-sized Planet", AJ, 155, 10 (2018)

[63] Johns-Krull, C.M., McLane, J.N., Prato, L., Crockett, C.J., Jaffe, D.T., Hartigan, P.M., Beichman, C.A., Mahmud, N.I., Chen, W., Skiff, B.A., Cauley, P.W., Jones, J.A., and Mace, G.N., "A Candidate Young Massive Planet in Orbit around the Classical T Tauri Star CI Tau", ApJ, 826, 206 (2016)

[64] Klein, B., Donati, J.-F., Moutou, C., Delfosse, X., Bonfils, X., Martioli, E., Fouque, P., Cloutier, R., Artigau, E., Doyon, R., Hebrard, G., Morin, J., Rameau, J., Plavchan, P., and Gaidos, E., "Investigating the young AU Mic system with SPIRou: large-scale stellar magnetic field and close-in planet mass", MNRAS, 502, 188-205 (2021)

[65] Suárez Mascareño, A., Damasso, M., Lodieu, N., Sozzetti, A., Béjar, V. J. S., Benatti, S., Zapatero Osorio, M. R., Micela, G., Rebolo, R., Desidera, S., Murgas, F., Claudi, R., González Hernández, J. I., Malavolta, L., del Burgo, C., D'Orazi, V., Amado, P. J., Locci, D., Tabernero, H. M., Marzari, F., Aguado, D. S., Turrini, D., Cardona Guillén, C., Toledo-Padrón, B., Maggio, A., Aceituno, J., Bauer, F. F., Caballero, J. A., Chinchilla, P., Esparza-Borges, E., González-Álvarez, E., Granzer, T., Luque, R., Martín, E. L., Nowak, G., Oshagh, M., Pallé, E., Parviainen, H., Quirrenbach, A., Reiners, A., Ribas, I., Strassmeier, K. G., Weber, M., and Mallonn, M., "Rapid contraction of giant planets orbiting the 20-million-year-old star V1298 Tau", Nature Astronomy, 6, 232-240 (2021)

[66] Mordasini, C., Alibert, Y., Klahr, H., and Henning, T., "Characterization of exoplanets from their formation. I. Models of combined planet formation and evolution", A&A, 547, 111 (2012)

[67] Mordasini, C., Alibert, Y., Georgy, C., Dittkrist, K.-M., Klahr, H., and Henning, T., "Characterization of exoplanets from their formation. II. The planetary mass-radius relationship", A&A, 547, 112 (2012)

[68] Mordasini, C., Alibert, Y., Benz, W., Klahr, H., and Henning, T., "Extrasolar planet population synthesis . IV. Correlations with disk metallicity, mass, and lifetime", A&A, 541, 97 (2012)

[69] Fortney, J.J., Marley, M.S., and Barnes, J.W., "Planetary Radii across Five Orders of Magnitude in Mass and Stellar Insolation: Application to Transits", ApJ, 659, 1661-1672 (2007)

[70] David, T.J., Contardo, G., Sandoval, A., Angus, R., Lu, Y.L., Bedell, M., Curtis, J.L., Foreman-Mackey, D., Fulton, B.J., Grunblatt, S.K., and Petigura, E.A., "Evolution of the Exoplanet Size Distribution: Forming Large Super-Earths Over Billions of Years", AJ, 161, 265 (2021)

[71] Pascucci, I., Mulders, G.D., and Lopez, E., "The Impact of Stripped Cores on the Frequency of Earth-size Planets in the Habitable Zone", ApJ, 883, 15 (2019)

[72] Kipping, D., Bryson, S., Burke, C., Christiansen, J., Hardegree-Ullman, K., Quarles, B., Hansen, B., Szulágyi, J., Teachey, A., "An exomoon survey of 70 cool giant exoplanets and the new candidate Kepler-1708 b-i", Nature Astronomy, 6, 367-380 (2022)

[73] Lazzoni, C., Zurlo, A., Desidera, S., Mesa, D., Fontanive, C., Bonavita, M., Ertel, S., Rice, K., Vigan, A., Boccaletti, A., Bonnefoy, M., Chauvin, G., Delorme, P., Gratton, R., Houllé, M., Maire, A. L., Meyer, M., Rickman, E., Spalding, E. A., Asensio-Torres, R., Langlois, M., Müller, A., Baudino, J. -L., Beuzit, J. -L., Biller, B., Brandner, W., Buenzli, E., Cantalloube, F., Cheetham, A., Cudel, M., Feldt, M., Galicher, R., Janson, M., Hagelberg, J., Henning, T., Kasper, M., Keppler, M., Lagrange, A. -M., Lannier, J., LeCoroller, H., Mouillet, D., Peretti, S., Perrot, C., Salter, G., Samland, M., Schmidt, T., Sissa, E., and Wildi, F., "The search for disks or planetary objects around directly imaged companions: a candidate around DH Tauri B", A&A, 614, 131 (2020)

[74] Reynolds, R.T., Squyres, S.W., Colburn, D.S., and McKay, C.P., "On the habitability of Europa", Icarus, 56, 246-254 (1983)

[75] Vanderburg, A., Rappaport, S.A., and Mayo, A.W., "Detecting Exomoons via Doppler Monitoring of Directly Imaged Exoplanets", AJ, 156, 184 (2018)

[76] Canup, R.M. and Ward, W.R., "A common mass scaling for satellite systems of gaseous planets", Nature, 441, 834-839 (2006)

[77] Ruffio, J.-B., Konopacky, Q.M., Barman, T., Macintosh, B., Hoch, K.K.W., De Rosa, R.J., Wang, J.J., Czekala, I., and Marois, C., "Deep Exploration of the Planets HR 8799 b, c, and d with Moderate-resolution Spectroscopy", AJ, 162, 290 (2021)

[78] Vanderburg, A. and Rodriguez, J.E., "First Doppler Limits on Binary Planets and Exomoons in the HR 8799 System", ApJ, 922, 2 (2021)

[79] Ruffio, J.-B., Horstman, K., Mawet, D., Rosenthal, L.J., Batygin, K., Wang, J.J., Millar-Blanchaer, M., Wang, J., Fulton, B.J., Konopacky, Q.M., Agrawal, S., Hirsch, L.A., Howard, A.W., Blunt, S., Nielsen, E., Baker, A., Bartos, R., Bond, C.Z., Calvin, B., Cetre, S., Delorme, J.-R., Doppmann, G., Echeverri, D., Finnerty, L., Fitzgerald, M.P., Jovanovic, N., López, R., Martin, E.C., Morris, E., Pezzato, J., Ruane, G., Sappey, B., Schofield, T., Skemer, A., Venenciano, T., Wallace, J.K.,Wallack, N.L., Wizinowich, P., and Xuan, J.W., "Detecting Exomoons from Radial Velocity Measurements of Self-luminous Planets: Application to Observations of HR 7672 B and Future Prospects", AJ, 165, 113 (2023)

[80] Batygin, K. and Morbidelli, A., "Formation of Giant Planet Satellites", ApJ, 894, 143 (2020)

[81] Feinstein, A.D., Schlieder, J.E., Livingston, J.H., Ciardi, D.R., Howard, A.W., Arnold, L., Barentsen, G., Bristow, M., Christiansen, J.L., Crossfield, I.J.M., Dressing, C.D., Gonzales, E.J., Kosiarek, M., Lintott, C.J., Miller, G., Morales, F.Y., Petigura, E.A., Thackeray, B., Ault, J., Baeten, E., Jonkeren, A.F., Langley, J., Moshinaly, H., Pearson, K., Tanner, C., and Treasure, J., "K2-288Bb: A Small Temperate Planet in a Low-mass Binary System Discovered by Citizen Scientists", AJ, 157, 40 (2019)

[82] Ciardi, D.R., Beichman, C.A., Horch, E.P., and Howell, S.B., "Understanding the Effects of Stellar Multiplicity on the Derived Planet Radii from Transit Surveys: Implications for Kepler, K2, and TESS", ApJ, 805, 16 (2015)

[83] Kraus, A.L., Ireland, M.J., Huber, D., Mann, A.W., and Dupuy, T.J., "The Impact of Stellar Multiplicity on Planetary Systems. I. The Ruinous Influence of Close Binary Companions", AJ, 152, 8 (2016)

[84] Moe, M. and Kratter, K.M., "Impact of binary stars on planet statistics - I. Planet occurrence rates and trends with stellar mass", MNRAS, 507, 3593-3611 (2021)

[85] Vos, J.M., Biller, B.A., Allers, K.N., Faherty, J.K., Liu, M.C., Metchev, S., Eriksson, S., Manjavacas, E., Dupuy, T.J., Janson, M., Radigan-Hoffman, J., Crossfield, I., Bonnefoy, M., Best, W.M.J., Homeier, D., Schlieder, J.E., Brandner, W., Henning, T., Bonavita, M., and Buenzli, E., "Spitzer Variability Properties of Low-gravity L Dwarfs", AJ, 160, 38 (2020)

[86] Millar-Blanchaer, M.A., Girard, J.H., Karalidi, T., Marley, M.S., van Holstein, R.G., Sengupta, S., Mawet, D., Kataria, T., Snik, F., de Boer, J., Jensen-Clem, R., Vigan, A., and Hinkley, S., "Detection of Polarization due to Cloud Bands in the Nearby Luhman 16 Brown Dwarf Binary", ApJ, 894, 42 (2020)

[87] Bailey, J., Kedziora-Chudczer, L., and Bott, K., "Polarized radiative transfer in planetary atmospheres and the polarization of exoplanets", MNRAS, 480, 1613-1625 (2018)

[88] Chakrabarty, A. and Sengupta, S., "Generic Models for Disk-resolved and Disk-integrated Phase-dependent Linear Polarization of Light Reflected from Exoplanets", ApJ, 917, 83 (2021)

[89] Madhusudhan, N. and Burrows, A., "Analytic Models for Albedos, Phase Curves, and Polarization of Reflected Light from Exoplanets", ApJ, 747, 25 (2012)

[90] Karalidi, T., Stam, D.M., and Guirado, D., "Flux and polarization signals of spatially inhomogeneous gaseous exoplanets", A&A, 555, 127 (2013)

[91] Rossi, L., Berzosa-Molina, J., Desert, J.-M., Fossati, L., Muñoz, A. G., Haswell, C., Kabath, P., Kislyakova, K., Stam, D., and Vidotto, A., "Spectropolarimetry as a tool for understanding the diversity of planetary atmospheres", Experimental Astronomy, 54, 1187-1196 (2022)

[92] Miles-Páez, P.A., Pallé, E., and Zapatero Osorio, M.R., "Simultaneous optical and near-infrared linear spectropolarimetry of the earthshine", A&A, 562, 5 (2014)

[93] Stam, D.M., "Spectropolarimetric signatures of Earth-like extrasolar planets", A&A, 482, 989-1007 (2008)

[94] Karalidi, T., Stam, D.M., and Hovenier, J.W., "Flux and polarisation spectra of water clouds on exoplanets", A&A, 530, 69 (2011)

[95] Gordon, K.E., Karalidi, T., Bott, K.M., Miles-Páez, P.A., Mulder, W., and Stam, D.M., "Polarized Signatures of a Habitable World: Comparing Models of an Exoplanet Earth with Visible and Near-infrared Earthshine Spectra", ApJ, 945, 166 (2023)

[96] Kochukhov, O., "Magnetic fields of M dwarfs", A&Arv, 29, 1 (2021)

[97] Wade, G.A., Donati, J.-F., Landstreet, J.D., and Shorlin, S.L.S., "Spectropolarimetric measurements of magnetic Ap and Bp stars in all four Stokes parameters", MNRAS, 313, 823-850 (2000)

[98] Donati, J. -F.; Howarth, I. D.; Jardine, M. M.; Petit, P.; Catala, C.; Landstreet, J. D.; Bouret, J. -C.; Alecian, E.; Barnes, J. R.; Forveille, T.; Paletou, F.; Manset, N., "The surprising magnetic topology of τ Sco: fossil remnant or dynamo output?", MNRAS, 370, 629-644 (2006)

[99] Kochukhov, O., Mantere, M.J., Hackman, T., and Ilyin, I., "Magnetic field topology of the RS CVn star II Pegasi", A&A, 550, 84 (2013)

[100] Rosén, L., Kochukhov, O., and Wade, G.A., "First Zeeman Doppler Imaging of a Cool Star Using all Four Stokes Parameters", ApJ, 805, 169 (2015)

[101] Klein, B., Donati, J.-F., Hébrard, E.M., Zaire, B., Folsom, C.P., Morin, J., Delfosse, X., and Bonfils, X., "The large-scale magnetic field of Proxima Centauri near activity maximum", MNRAS, 500, 1844-1850 (2021)

[102] Williamo, T., Hackman, T., Lehtinen, J.J., Korpi-Lagg, M., and Kochukhov, O., "V889 Her: abrupt changes in the magnetic field or differential rotation?", OJAp, 5, 10 (2022)

[103] Khodachenko, M.L., Ribas, I., Lammer, H., Grießmeier, J.-M., Leitner, M., Selsis, F., Eiroa, C., Hanslmeier, A., Biernat, H.K., Farrugia, C.J., and Rucker, H.O., "Coronal Mass Ejection (CME) Activity of Low Mass M Stars as An Important Factor for The Habitability of Terrestrial Exoplanets. I. CME Impact on Expected Magnetospheres of Earth-Like Exoplanets in Close-In Habitable Zones", Astrobiology, 7, 167 (2007)

[104] Luger, R. and Barnes, R., "Extreme Water Loss and Abiotic $O_2$ Buildup on Planets Throughout the Habitable Zones of M Dwarfs", Astrobiology, 15, 119-143 (2015)

[105] Kislyakova, K. G., Noack, L., Johnstone, C. P., Zaitsev, V. V., Fossati, L., Lammer, H., Khodachenko, M. L., Odert, P., and Güdel, M., "Magma oceans and enhanced volcanism on TRAPPIST-1 planets due to induction heating", Nature Astronomy, 1, 878 (2017)

[106] Ridgway, R. J., Zamyatina, M., Mayne, N. J., Manners, J., Lambert, F. H., Braam, M., Drummond, B., Hébrard, E., Palmer, P. I., and Kohary, K., "3D modelling of the impact of stellar activity on tidally locked terrestrial exoplanets: atmospheric composition and habitability", MNRAS, 518, 2472-2496 (2023)

[107] Browning, M.K., Weber, M.A., Chabrier, G., and Massey, A.P., "Theoretical Limits on Magnetic Field Strengths in Low-mass Stars", ApJ, 818, 189 (2016)

[108]  Khurana, K.K., Kivelson, M.G., Vasyliunas, V.M., Krupp, N., Woch, J., Lagg, A., Mauk, B.H., and Kurth, W.S., "The configuration of Jupiter's magnetosphere", Jupiter. The Planet, Satellites and Magnetosphere, Ed. F. Bagenal, T.E. Dowling, W.B. McKinnon, Cambridge University Press, 593-616 (2004)

[109]  Kirkpatick, J.D., "New Spectral Types L and T", ARA&A, 43, 195-245 (2005)

[110]  Hsu, C.-C., Burgasser, A.J., Theissen, C.A., Gelino, C.R., Birky, J.L., Diamant, S.J.M., Bardalez Gagliuffi, D.C., Aganze, C., Blake, C.H., and Faherty, J d.K., "The Brown Dwarf Kinematics Project (BDKP). V. Radial and Rotational Velocities of T Dwarfs from Keck/NIRSPEC High-resolution Spectroscopy", ApJS, 257, 45 (2021)

[111]  Gagné, J., Mamajek, E.E., Malo, L., Riedel, A., Rodriguez, D., Lafrenière, D., Faherty, J.K., Roy-Loubier, O., Pueyo, L., Robin, A.C., and Doyon, R., "BANYAN. XI. The BANYAN Σ Multivariate Bayesian Algorithm to Identify Members of Young Associations with 150 pc", ApJ, 856, 23 (2018)

[112]  Mamajek, E.E., Barenfeld, S.A., Ivanov, V.D., Kniazev, A.Y., Väisänen, P., Beletsky, Y., and Boffin, H.M.J., "The Closest Known Flyby of a Star to the Solar System", ApJ, 800, 17 (2015)

[113]  Burgasser, A., Apai, D., Bardalez Gagliuffi, D., Blake, C., Gagne, J., Konopacky, Q., Martin, E., Metchev, S., Reiners, A., Schlawin, E., Sousa-Silva, C., and Vos, J., "High-Resolution Spectroscopic Surveys of Ultracool Dwarf Stars & Brown Dwarfs", BAAS Astro2020 White Papers, 51, 547 (2019)

[114]  Konopacky, Q.M., Ghez, A.M., Barman, T.S., Rice, E.L., Bailey, J.I., III, White, R.J., McLean, I.S., and Duchêne, G., "High-precision Dynamical Masses of Very Low Mass Binaries", ApJ, 711, 1087 (2010)

[115]  Burgasser, A.J., Cruz, K.L., Cushing, M., Gelino, C.R., Looper, D.L., Faherty, J.K., Kirkpatrick, J.D., and Reid, I.N., "SpeX Spectroscopy of Unresolved Very Low Mass Binaries. I. Identification of 17 Candidate Binaries Straddling the L Dwarf/T Dwarf Transition", ApJ, 710, 1142 (2010)

[116]  Bardalez Gagliuffi, D.C., Burgasser, A.J., Gelino, C.R., Looper, D.L., Nicholls, C.P., Schmidt, S.J., Cruz, K., West, A.A., Gizis, J.E., and Metchev, S., "SpeX Spectroscopy of Unresolved Very Low Mass Binaries. II. Identification of 14 Candidate Binaries with Late-M/Early-L and T Dwarf Components", ApJ, 794, 143 (2014)

[117]  Irwin, J., Berta, Z.K., Burke, C.J., Charbonneau, D., Nutzman, P., West, A.A., and Falco, E.E., "On the Angular Momentum Evolution of Fully Convective Stars: Rotation Periods for Field M-dwarfs from the MEarth Transit Survey", ApJ, 727, 56 (2011)

[118]  Ricker, G.R., Winn, J.N., Vanderspek, R., Latham, D.W., Bakos, G.Á., Bean, Ja.L., Berta-Thompson, Z.K., Brown, T.M., Buchhave, L., Butler, N.R., Butler, R.P., Chaplin, W.J., Charbonneau, D., Christensen-Dalsgaard, J., Clampin, M., Deming, D., Doty, J., De Lee, N., Dressing, C., Dunham, E.W., Endl, M., Fressin, F., Ge, J., Henning, T., Holman, M.J., Howard, A.W., Ida, S., Jenkins, J., Jernigan, G., Johnson, J.A., Kaltenegger, L., Kawai, N., Kjeldsen, H., Laughlin, G., Levine, A.M., Lin, D., Lissauer, J.J., MacQueen, P., Marcy, G., McCullough, P.R., Morton, T.D., Narita, N., Paegert, M., Palle, E., Pepe, F., Pepper, J., Quirrenbach, A., Rinehart, S.A., Sasselov, D., Sato, B., Seager, S., Sozzetti, A., Stassun, K.G., Sullivan, P., Szentgyorgyi, A., Torres, G., Udry, S., and Villasenor, J., "Transiting Exoplanet Survey Satellite (TESS)", JATIS, 1, 014003 (2015)

[119]  Delrez, L., Gillon, M., Queloz, D., Demory, B.-O., Almleaky, Y., de Wit, J., Jehin, E.., Triaud, A.H.M.J., Barkaoui, K., Burdanov, A., Burgasser, A.J., Ducrot, E., McCormac, J., Murray, C., Silva Fernandes, C., Sohy, S., Thompson, S.J., Van Grootel, V., Alonso, R., Benkhaldoun, Z., and Rebolo, Rafael, "SPECULOOS: a network of robotic telescopes to hunt for terrestrial planets around the nearest ultracool dwarfs", Proc. SPIE, 10700 (2018)

[120]  Tamburo, P. and Muirhead, P.S., "Design Considerations for a Ground-based Search for Transiting Planets around L and T Dwarfs", PASP, 131, 114401 (2019)

[121]  Metchev, S.A., Heinze, A., Apai, D., Flateau, D., Radigan, J., Burgasser, A., Marley, M.S., Artigau, É., Plavchan, P., and Goldman, B., "Weather on Other Worlds. II. Survey Results: Spots are Ubiquitous on L and T Dwarfs", ApJ, 799, 154 (2015)

[122]  Vos, J.M., Allers, K.N., and Biller, B.A., "The Viewing Geometry of Brown Dwarfs Influences Their Observed Colors and Variability Amplitudes", ApJ, 842, 78 (2017)

[123]  Reiners, A. and Basri, G., "Measuring Magnetic Fields in Ultracool Stars and Brown Dwarfs", ApJ, 644, 497-509 (2006)

[124]  Terrien, R.C., Keen, A., Oda, K., Parts, W., Stefánsson, G., Mahadevan, S., Robertson, P., Ninan, J.P., Beard, C., Bender, C.F., Cochran, W.D., Cunha, K., Diddams, S.A., Fredrick, C., Halverson, S., Hearty, F., Ickler, A., Kanodia, S., Libby-Roberts, J.E., Lubin, J., Metcalf, A.J., Olsen, F., Ramsey, L.W., Roy, A., Schwab, C., Smith, V.V., and Turner, B., "Rotational Modulation of Spectroscopic Zeeman Signatures in Low-mass Stars", ApJ, 927, 11 (2022)

[125]  Donati, J.-F. and Brown, S.F., "Zeeman-Doppler imaging of active stars. V. Sensitivity of maximum entropy magnetic maps to field orientation", A&A, 326, 1135-1142 (1997)

[126] Morin, J., Bouret, J.-C., Neiner, C., Aerts, C., Bagnulo, S., Catala, C., Charbonnel, C., Evans, C., Fossati, L., Garcia, M., Gómez de Castro, A.I., Herrero, A., Hussain, G., Kaper, L., Kochukhov, O., Konstantinova-Antova, R., de Koter, A., Kraus, M., Jiříkrtička, L., Agnes, L., Theresa, M., Georges, P., Pascal, S., Steve, S., Sami, S., Beate, S., Antoine, V., Aline, V., and Jorick S, "Stellar Physics with High-Resolution UV Spectropolarimetry", ESA Voyage 2050 White Paper, arXiv:1908.01545 (2019)

[127] Browning, M.K., "Simulations of Dynamo Action in Fully Convective Stars", ApJ, 676, 1262-1280 (2008)

[128] Brown, B.P., Oishi, J. S., Vasil, G.M., Lecoanet, D., and Burns, K. J., "Single-hemisphere Dynamos in M-dwarf Stars", ApJ, 902, 3 (2020)

[129] Charnay, B., Bézard, B., Baudino, J. -L., Bonnefoy, M., Boccaletti, A., and Galicher, R., "A Self-consistent Cloud Model for Brown Dwarfs and Young Giant Exoplanets: Comparison with Photometric and Spectroscopic Observations", ApJ, 854, 172 (2018)

[130] Buenzli, E., Apai, D., Morley, C.V., Flateau, D., Showman, A.P., Burrows, A., Marley, M.S., Lewis, N.K., and Reid, I.N., "Vertical Atmospheric Structure in a Variable Brown Dwarf: Pressure-dependent Phase Shifts in Simultaneous Hubble Space Telescope-Spitzer Light Curves", ApJ, 760, 31 (2012)

[131] Apai, D., Radigan, J., Buenzli, E., Burrows, A., Reid, I.N., and Jayawardhana, R., "HST Spectral Mapping of L/T Transition Brown Dwarfs Reveals Cloud Thickness Variations", ApJ, 768, 121 (2013)

[132] Marley, M.S. and Robinson, T.D., "On the Cool Side: Modeling the Atmospheres of Brown Dwarfs and Giant Planets", ARA&A, 53, 279-323 (2015)

[133] GRAVITY Collaboration, Abuter, R., Amorim, A., Bauböck, M., Berger, J. P., Bonnet, H., Brandner, W., Cardoso, V., Clénet, Y., de Zeeuw, P. T., Dexter, J., Eckart, A., Eisenhauer, F., Förster Schreiber, N. M., Garcia, P., Gao, F., Gendron, E., Genzel, R., Gillessen, S., Habibi, M., Haubois, X., Henning, T., Hippler, S., Horrobin, M., Jiménez-Rosales, A., Jochum, L., Jocou, L., Kaufer, A., Kervella, P., Lacour, S., Lapeyrère, V., Le Bouquin, J. -B., Léna, P., Nowak, M., Ott, T., Paumard, T., Perraut, K., Perrin, G., Pfuhl, O., Rodríguez-Coira, G., Shangguan, J., Scheithauer, S., Stadler, J., Straub, O., Straubmeier, C., Sturm, E., Tacconi, L. J., Vincent, F., von Fellenberg, S., Waisberg, I., Widmann, F., Wieprecht, E., Wiezorrek, E., Woillez, J., Yazici, S. and Zins, G., "Detection of the Schwarzschild precession in the orbit of the star S2 near the Galactic centre massive black hole", A&A, 636, 5 (2020)

[134] Safronova, M.S., "The Search for Variation of Fundamental Constants with Clocks", Annalen der Physik, 531, 1800364 (2019)

[135] Webb, J.K., Flambaum, V.V., Churchill, C.W., Drinkwater, M.J., and Barrow, J.D., "Search for Time Variation of the Fine Structure Constant", Phys. Rev. Lett., 82, 884–887 (1999)

[136] Hu, J., Webb, J. K., Ayres, T. R., Bainbridge, M. B., Barrow, J. D., Barstow, M. A., Berengut, J. C., Carswell, R. F., Dumont, V., Dzuba, V., Flambaum, V. V., Lee, C. C., Reindl, N., Preval, S. P., and Tchang-Brillet, W.Ü.L., "Measuring the fine structure constant on a white dwarf surface; a detailed analysis of Fe V absorption in G191-B2B", MNRAS, (2021)

[137] Rahmani, H., Maheshwari, N., and Srianand. R., "Constraining the variation in the fine-structure constant using SDSS DR7 quasi-stellar object spectra", MNRAS, 439, L70–L74 (2014)

[138] Hees, A., Do, T., Roberts, B. M., Ghez, A. M., Nishiyama, S., Bentley, R. O., Gautam, A. K., Jia, S., Kara, T., Lu, J. R., Saida, H., Sakai, S., Takahashi, M., and Takamori, Y. "Search for a Variation of the Fine Structure Constant around the Supermassive Black Hole in Our Galactic Center", Phys. Rev. Lett., 124, 081101 (2020)

[139] Genzel, R., Schödel, R., Ott, T., Eckart, A., Alexander, T., Lacombe, F., Rouan, D., and Aschenbach, B. "Near-infrared flares from accreting gas around the supermassive black hole at the Galactic Centre", Nature, 425, 934–937, (2003)

[140] Fragione, G. and Loeb, A., "An Upper Limit on the Spin of SgrA* Based on Stellar Orbits in Its Vicinity", ApJ, 901, L32 (2020)

[141] Will, C.M. "Testing the General Relativistic "No-Hair" Theorems Using the Galactic Center Black Hole Sagittarius A*", ApJ, 674, L25 (2008)

[142] Gould, M., Vincent, F. H., Paumard, T., and Perrin, G., "General relativistic effects on the orbit of the S2 star with GRAVITY", A&A, 608, A60 (2017)

[143] Tsuboi, M., Kitamura, Y., Tsutsumi, T., Uehara, K., Miyoshi, M., Miyawaki, R., and Miyazaki, A., "The Second Galactic Center Black Hole? A Possible Detection of Ionized Gas Orbiting around an IMBH Embedded in the Galactic Center IRS13E Complex", ApJ, 850, L5 (2017)